# Imaging nanoscale photocarrier traps in solar water-splitting catalysts

Levi D. Palmer,[1] Wonseok Lee,[1] Pushp Raj Prasad,[2] Bradley W. Layne,[2] Han-Hsuan Wu,[3] Zejie Chen,[2] Jianguo Wen,[4] Yuzi Liu,[4] Xiaoqing Pan,[3,5,6] A. Alec Talin,[7] Akihiko Kudo,[8] Shane Ardo,[2,6,9] Joseph P. Patterson,[2,6] Thomas E. Gage,[4]* Scott K. Cushing[1]*

**Affiliations:**

[1]Division of Chemistry and Chemical Engineering, California Institute of Technology; Pasadena, CA 91125, USA

[2]Department of Chemistry, University of California, Irvine; Irvine, CA 92697, USA

[3]Department of Physics and Astronomy, University of California, Irvine; Irvine, CA 92697, USA

[4]Center for Nanoscale Materials, Argonne National Laboratory; Lemont, IL 60439, USA

[5]Irvine Materials Research Institute (IMRI), University of California, Irvine; Irvine, CA 92697, USA

[6]Department of Materials Science and Engineering, University of California, Irvine; Irvine, CA 92697, USA

[7]Sandia National Laboratories; Livermore, CA 94550, USA

[8]Department of Applied Chemistry, Faculty of Science, Tokyo University of Science; Tokyo 162-8601, Japan

[9]Department of Chemical and Biomolecular Engineering, University of California, Irvine; Irvine, CA 92697, USA

*Corresponding author. Email: tgage@anl.gov (T.E.G.)

*Corresponding author. Email: scushing@caltech.edu (S.K.C.)

**Abstract:** Defects trap photocarriers and hinder solar water splitting. The nanoscale photocarrier transport, trapping, and recombination mechanisms are usually inferred from ensemble-averaged measurements and remain elusive. Because an individual high-performing nanoparticle photocatalyst may outperform the ensemble average, design rules that would otherwise enhance catalytic efficiency remain unclear. Here, we introduce photomodulated electron energy-loss spectroscopy (EELS) in an optically coupled scanning transmission electron microscope (STEM) to map photocarrier localization. Using rhodium-doped strontium titanate ($SrTiO_3$:Rh) solar water-splitting nanoparticles, we directly image the carrier densities concentrated at oxygen-vacancy surface trap states. This is achieved by separating photothermal heating from photocarrier populations through experimental and computational analyses of low-loss spectra. Photomodulated STEM-EELS enables angstrom-scale imaging of defect-induced photocarrier traps and their impact on photocatalytic efficiency.

**Main Text:**

Efficient solar water splitting requires long photocarrier lifetimes to ensure that carriers can sufficiently transport to catalytic sites (*1*). Nanoparticle photocatalysts are often engineered to suppress carrier recombination and to spatially separate electron–hole pairs. Common strategies to increase photocarrier lifetimes are cocatalysts that separate charge carriers, doping strategies that suppress trap sites, and flux treatments that increase crystallinity, enabling nearly 100% internal quantum efficiency for one select catalyst composition (*2*, *3*). However, most photocatalysts remain plagued by photocarrier traps and recombination sites (*4*, *5*).

Although these trap and recombination sites drive carrier recombination and impede carrier transport, they have not been directly imaged under steady-state illumination conditions relevant to solar water splitting. Steady-state illumination results in photothermal and photocarrier effects that must be separately resolved. Stroboscopic scattering microscopy accurately images photocarriers and temperature down to the micrometer-scale optical diffraction limit (*6*). Ultrafast scanning probe microscopy images photocarrier dynamics with 1–100 nm resolution, but its contrast mechanism provides only indirect information about bulk carrier dynamics and is insensitive to heat (*7*, *8*). Electron energy-loss spectroscopy (EELS) thermometry in the scanning transmission electron microscope (STEM) directly images angstrom-scale temperatures from femtoseconds to seconds (*9*–*12*). STEM-EELS has yet to resolve excited-state photocarriers on the same length scale.

We develop photomodulated STEM-EELS in an optically coupled STEM to directly image steady-state photocarriers in 1 wt% rhodium-doped strontium titanate ($SrTiO_3$:Rh) nanoparticles with angstrom-scale resolution. We first use core-loss STEM-EELS and atomic-resolution STEM imaging to identify carrier trap sites, including surface oxygen vacancies and copper cocatalysts. We then image photoexcited carriers in these traps with photomodulated STEM-EELS of the low-loss spectrum. The loss function measured by EELS is traditionally elusive, especially for metal oxide insulators that do not follow the free electron model, resulting in overlapping bulk plasmon and interband transition peaks. We perform time-dependent density functional theory (TDDFT) calculations to determine which VB electrons contribute to each peak of the loss function. As a result, we distinguish how heat and photocarriers separately influence select low-loss peaks, supported by experimental heating control experiments. We image trapped photocarriers at the $SrTiO_3$:Rh nanoparticle's surface at a concentration roughly 70% higher than the nanoparticle's bulk. Overall, we use photomodulated STEM-EELS imaging to visualize nanoscale carrier trapping in nanoparticle photocatalysts—directly imaging carrier trapping phenomena previously inferred by ensemble-averaged measurements. We demonstrate how photomodulated STEM-EELS imaging can be used to resolve carriers and heat in nanostructured photocatalysts, optoelectronics, and quantum photonics.

**Platform for imaging photocarrier trap states**

We perform photomodulated STEM-EELS in an optically coupled STEM equipped with a continuous-wave laser (Fig. 1A). The 3.1 eV laser induces photocarriers by photoexciting $SrTiO_3$:Rh nanoparticles above their 2.8 eV bandgap (*13*). A resistive heater is used for temperature-dependent control measurements. Simultaneous EEL spectral acquisition is performed with angstrom-scale precision. EELS acquisition is achieved using a prism and a direct electron detector. This combined platform enables photomodulated STEM-EELS imaging, thereby resolving photocarrier trapping at angstrom-to-micrometer structural features that govern solar water splitting efficiencies (Fig. 1B). Photographs of the experimental setup are shown in Fig. S1.

Prior studies report that $SrTiO_3$:Rh nanoparticles' photocarriers are (1) spatially separated by cocatalysts, (2) trapped by oxygen vacancies, and (3) localized to rhodium dopants (Fig. 1B) (*3*, *13*–*15*). $SrTiO_3$ must be doped to enable visible light absorption. However, these dopants often create defect states that restrict photocarrier mobilities and lifetimes. For example, rhodium dopants introduce occupied and unoccupied Rh *4d* states above the O *2p* valence band (VB) maximum of undoped $SrTiO_3$ (Fig. 1C). Visible light photoexcitation of $SrTiO_3$:Rh promotes electrons from the Rh *4d* VB states into the Ti *3d* conduction band (CB). These rhodium-localized midgap states, as well as oxygen vacancies, create recombination and trap sites, posing an engineering challenge.

Photomodulated STEM-EELS resolves excited-state photocarrier localization both spatially (Fig. 1B) and energetically (Fig. 1C). For photomodulated STEM-EELS, both low-loss and core-loss spectra are measurable, each with distinct advantages. While element-specific core-level excitations are proven to distinguish carrier and heat signatures (*16*–*18*), low-loss EELS yields orders-of-magnitude more signal and directly measure free carrier density. However, unlike core-loss spectroscopy, photothermal and photoexcited carrier effects have not been separated for low-loss EELS. We therefore employ TDDFT calculations to project select low-loss peaks onto the $SrTiO_3$ band structure. This spectrally resolved approach allows us to interpret the origin of the electronic structure of each low-loss EELS peak and elucidate how each mode is influenced by either photothermal heating or photoexcited carriers.

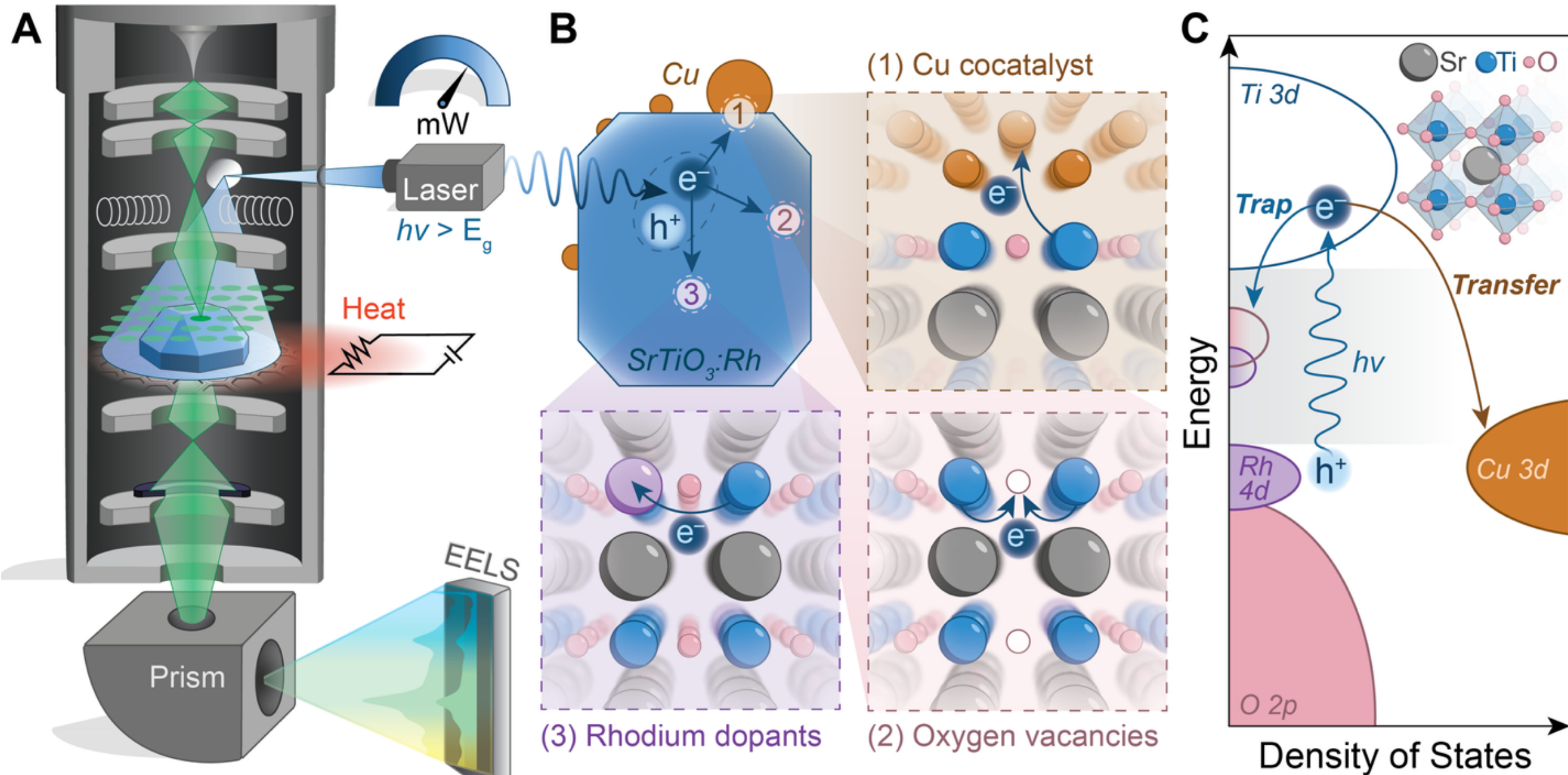

**Fig. 1. Photomodulated STEM-EELS images excited-state photocarrier localization**. (**A**) Schematic of the photomodulated STEM-EELS platform, in which continuous-wave laser illumination is coupled into the electron microscope. A resistive heater enables control experiment to differentiate photothermal and photocarrier effects. (**B**) Photocarrier trapping mechanisms in a $SrTiO_3$:Rh nanoparticle. Photogenerated electrons ($e^-$) and holes ($h^+$) may (1)

transfer to cocatalysts, (2) trap at oxygen vacancies, or (3) localize at rhodium dopants. (**C**) An energetic landscape of photocarrier relaxation pathways in $SrTiO_3$:Rh. The $SrTiO_3$ unit cell is inset. Trapping at midgap Rh *4d* and oxygen-vacancy states (shaded), along with transfer to the Cu cocatalyst, constitute primary decay channels for photoexcited electrons. Photomodulated STEM-EELS images these localization processes with angstrom-scale spatial and sub-eV spectral sensitivities.

## Ground-state STEM characterization

We characterize the ground-state properties of potential electron traps in a $SrTiO_3$:Rh nanoparticle (Fig. 2A). Fig. S2 shows an annular dark-field (ADF) images of the nanoparticle without false coloration. Our prior work investigated the location of rhodium dopants at atomic resolution, demonstrating that rhodium can substitute for both strontium and titanium atoms in the $SrTiO_3$ lattice, with a preference for titanium substitution and a uniform dopant distribution across the nanoparticle (*19*). Meanwhile, oxygen vacancies in $SrTiO_3$ nanoparticles are localized at the surface (*20*–*22*). Oxygen vacancies are thermodynamically favored defects in $SrTiO_3$. These well-studied vacancies form readily under high-temperature synthesis conditions and are further promoted by aliovalent Rh dopants and excess Sr used in our synthetic approach (*15*, *23*, *24*).

We characterize the $SrTiO_3$:Rh surface oxygen vacancies using core-loss STEM-EELS (Fig. 2B). Oxygen vacancies act as n-type dopants in $SrTiO_3$ while increasing polaron formation, introducing trap states, and ultimately limiting carrier mobility (*14*, *24*, *25*). Core-loss EELS probes core electrons transitioning into the $SrTiO_3$:Rh CBs, relaying the nanoparticle's electronic structure and composition with elemental specificity (*26*). We measure the Ti $L_{2,3}$ and O K edges of $SrTiO_3$:Rh with 2 nm resolution (Fig. 2B). Comparing spectra averaged at the nanoparticle's surface and bulk indicates that oxygen vacancies broaden the Ti $L_{2,3}$ edge and decrease the O K edge signal (*27*, *28*). We employ DFT and the Bethe–Salpeter equation (BSE) (DFT+BSE) to model core-loss spectra of bulk $SrTiO_3$, shown as the overlaid solid lines in Fig. 2B (*29*, *30*). These undoped $SrTiO_3$ calculations are consistent with $SrTiO_3$:Rh, given the negligible 1 wt% rhodium dopant concentration.

Cocatalysts on doped $SrTiO_3$ spatially isolate photoexcited electrons from holes, increasing the photocarrier lifetime by minimizing electron–hole recombination (*3*, *31*, *32*). For a Cu–$SrTiO_3$:Rh junction, photoexcited electrons are expected to transfer into copper due to the heterojunction's Fermi energy alignment (*33*). We use core-loss STEM-EELS to characterize a copper nanoparticle on the $SrTiO_3$:Rh surface (Fig. 2C). We measure the nanoparticle as 80% Cu and 20% $Cu_2O$, determined using a non-negative least squares linear combination fit of Cu(0) metal and $Cu_2O$ oxide DFT+BSE-calculated spectra (*34*, *35*). Spatially-averaged STEM-EEL spectra further indicate 100% metallic Cu at the Cu–$SrTiO_3$:Rh interface, likely due to preferential oxidation along the nanoparticle's facets or interfacial electron accumulation suppressing oxidation (*36*, *37*). Nanoprobe diffraction measurements show that both the Cu and $SrTiO_3$ nanoparticles are [110]-oriented. Spectral DFT+BSE calculations and fitting, as well as diffraction data, are presented in the Supplementary Text and Figs. S3 and S4.

We perform atomic-resolution integrated differential phase contrast (iDPC) STEM imaging of the $SrTiO_3$:Rh surface atoms (Fig. 2D). iDPC simultaneously resolves the atomic positions of both strontium and titanium atomic columns, oxygen columns yield weaker contrast and are not resolved. We map the Sr and Ti atomic columns by identifying circular atomic features via template matching and fit their position with a 2D Gaussian profile. We analyze these fit positions to calculate the interatomic distances between Sr–Ti and Ti–Ti / Sr–Sr (Fig. 2E). The interatomic distances both increase by >0.5 Å at the nanoparticle's surface and indicate a ~1–2 nm oxygen-vacancy layer. Our iDPC image analysis is consistent with prior findings that increasing tensile strain in $SrTiO_3$ correlates with oxygen vacancy concentration, as evidenced by larger lattice parameters in oxygen-depleted $SrTiO_{3-x}$ (*38*). We attribute the bright contrast of the surface layer to the high oxygen vacancy concentration and the misorientation of atoms off zone axis. Atomic-resolution image fitting and analysis are shown in Figs. S5 and S6.

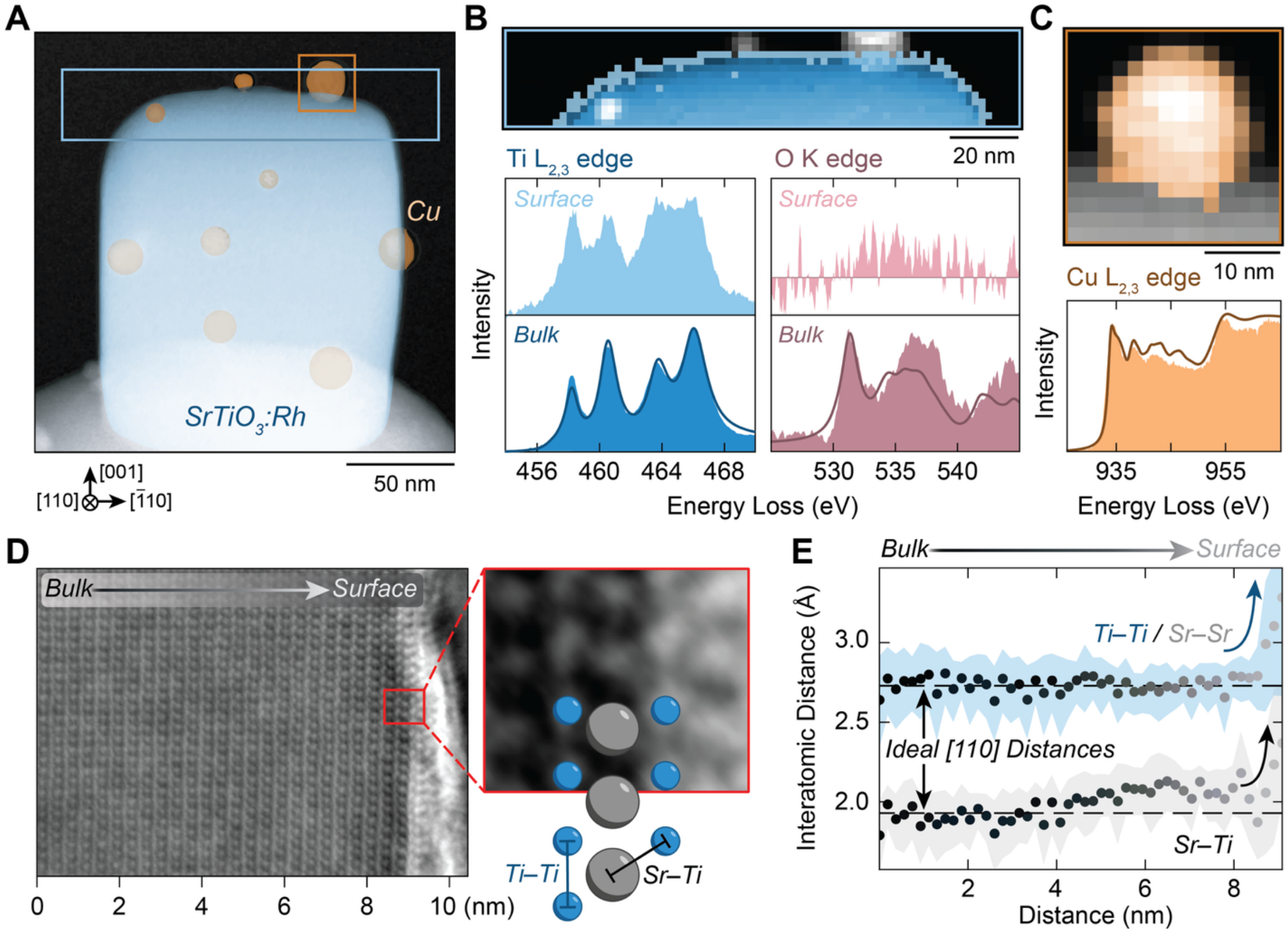

**Fig. 2. Ground-state imaging of photocarrier trap sites in $SrTiO_3$:Rh.** (**A**) False-colored ADF image of a $SrTiO_3$:Rh/Cu nanoparticle on sintered $SrTiO_3$:Rh. The STEM-EELS acquisition regions are outlined; $SrTiO_3$:Rh is shown in blue and Cu in orange. (**B**) Spatially resolved STEM-EELS map and corresponding Ti $L_{2,3}$- and O K-edge spectra from the $SrTiO_3$:Rh nanoparticle. DFT+BSE-calculated spectra of both edges for crystalline $SrTiO_3$ are overlaid. (**C**) STEM-EELS map of the Cu $L_{2,3}$ edge, indicating a Cu (80%) and $Cu_2O$ (20%) blend via DFT+BSE calculations. (**D**) Atomic-resolution iDPC-STEM image of an undoped $SrTiO_3$ nanoparticle, highlighting the oxygen-deficient $SrTiO_{3-x}$ surface. Sr and Ti atomic column positions are fitted to quantify deviations in lattice spacing as a function of depth. (**E**) Extracted Ti–Ti/Sr–Sr and Sr–Ti interatomic distances indicate tensile strain at the nanoparticle surface, consistent with strain induced by oxygen vacancies. Shading denotes the standard deviation of the extracted interatomic distance. All nanoparticles are imaged along the [110] zone axis.

## Low-loss EELS modeling

Photomodulated STEM-EELS readily resolves a material's loss function with low-loss EELS, as the signal-to-noise ratio is much higher than core-loss excitations. However, the loss function is complicated by overlapping interband transitions and bulk plasmon oscillations, and it remains poorly understood for insulators. Although the free electron model readily predicts metals' bulk plasmon energies, multi-element insulators like $SrTiO_3$ host complex interband transitions and plasmon modes confined to the VB (*39*). Bulk plasmons in $SrTiO_3$ are quantized oscillations of VB electrons, and prior work utilized vacuum ultraviolet reflectivity to label the interband transitions in $SrTiO_3$ (*40*, *41*). Interpreting changes in the loss function in situ during heating or photoexcitation introduces further complexity. However, the spectral effects induced by locally trapped photocarriers as opposed to photothermal heating must be distinguished. We implement band-structure-projected DFT and TDDFT calculations to interpret the origin of $SrTiO_3$'s loss function in both the ground and excited states (Fig. 3).

We first use DFT to orbitally resolve the band structure of $SrTiO_3$ (Fig. 3A). We project the atomic orbitals of $SrTiO_3$ into energy and momentum space. The calculation indicates that the VB and CB are composed of O *2p* and Ti *3d* orbitals, respectively. The Fermi level ($E_F$) is defined here as the VB maximum, where photoexcited holes thermalize, and photoexcited electrons thermalize to the CB minimum. The lower energy Sr *4p*, O *2s*, and Ti *3p* core-level orbitals will not host photocarrier dynamics. While our pseudopotentials underestimate the $SrTiO_3$ bandgap due to the commonly reported self-interaction error and derivative discontinuity, our band structure and TDDFT EELS calculations nevertheless align with the experiment (*42*, *43*), indicating that an insulator's band gap minimally influences its loss function. Figure S7 shows the calculated band structure without orbitals projected.

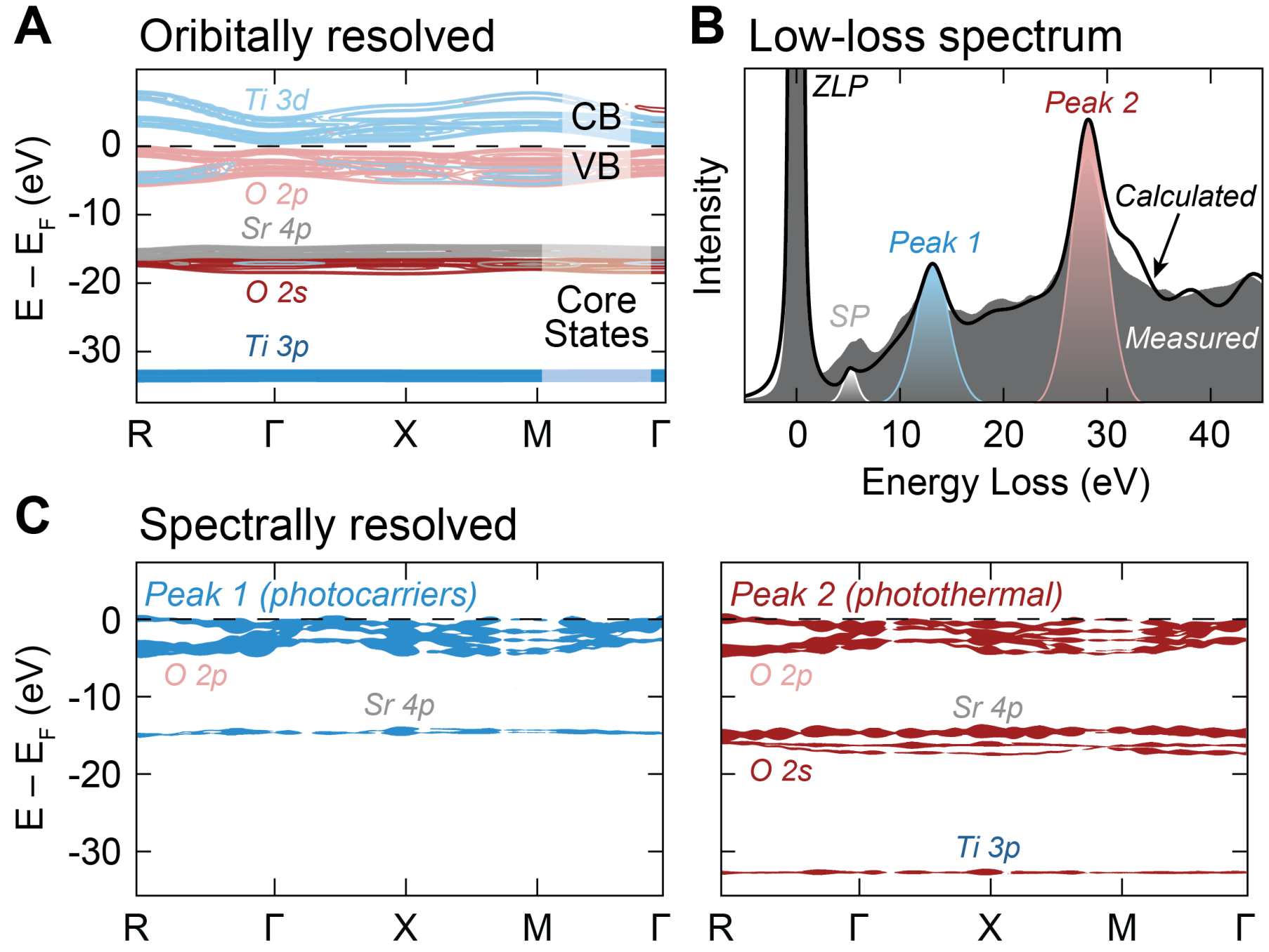

**Fig. 3. Identifying the origin of $SrTiO_3$ low-loss EELS peaks.** (**A**) Orbitally resolved band structure of $SrTiO_3$ calculated by DFT. Sr, Ti, and O orbitals are mapped onto the $SrTiO_3$ band structure with the band energy calibrated to $E_F$. (**B**) Measured low-loss EELS from a $SrTiO_3$

nanoparticle with the TDDFT-calculated spectrum overlaid as the solid black line. Spectra are aligned using the zero-loss peak (ZLP). One surface plasmon (SP) and two bulk low-loss peaks are identified for subsequent analysis. (**C**) Spectrally resolved band structure projections for peaks 1 and 2 in (B) are calculated using TDDFT. Peak 1 predominantly couples to O *2p* electrons near the valence band maximum, whereas peak 2 is produced by plasmon oscillations and interband transitions from valence electrons in all orbitals.

We use TDDFT to accurately interpret the low-loss spectrum of undoped $SrTiO_3$ (Fig. 3B). Low-loss EEL spectra are aligned by setting the zero-loss peak (ZLP) to 0 eV. The ZLP consists of elastically scattered electrons with a 0.45 eV full width at half maximum bandwidth. For $SrTiO_3$, numerous plasmon modes and interband transitions are resolved. This includes the first plasmon mode at 5 eV, which corresponds to surface plasmon (SP) modes of electrons at the valence band maximum (Fig. S8). Two intense bulk low-loss modes labeled "peak 1" and "peak 2" are highlighted. Both peaks exhibit multiple shoulder peaks. A TDDFT-calculated spectrum is overlaid on the experimental measurement in Fig. 3B. The TDDFT spectrum is calculated using the open-source Liouville-Lanczos solver in the turboEELS package (see Materials and Methods) (*44*, *45*).

We next determine which electrons in $SrTiO_3$ correspond to each peak (Fig. 3C). We modify the turboEELS solver to project select peaks of the loss function onto a material's band structure, thereby identifying the valence electrons that produce each peak. The developed spectral projection method follows our previously reported work and that of others (*46–48*), performed by storing the momentum-resolved electronic structure's probability density at select loss function energies (*49*, *50*). As shown in Fig. 3C, peak 1 corresponds to higher-lying valence states, while additional core states contribute to peak 2. As a result, peak 1 is highly sensitive to photoexcited carriers in the valence states. Peak 2, by contrast, is delocalized across multiple atoms and core states, making it more sensitive to lattice expansion following photothermal heating and far less sensitive to localized photocarriers. The SP is projected and spectrally resolved in Fig. S9, and Movie S1 depicts iterative spectral projections from 2–50 eV.

**Photomodulated STEM-EELS imaging**

Photomodulated STEM-EELS enables direct imaging of trapped photocarriers in $SrTiO_3$:Rh by spectrally resolving peak 1 (primarily sensitive to photocarriers) and 2 (probing photothermal heating as a control) as a function of laser illumination power. While all modes in the spectrum are sensitive to both photocarriers and heating, our spectral assignment differentiates photothermal temperature from photocarriers. Continuous-wave photoexcitation of the $SrTiO_3$:Rh nanoparticle induces subtle shifts in its average EEL spectrum acquired across a STEM-EELS image (Fig. 4A). Temperature-dependent measurements using a resistive heating holder are used as a control for photothermal heating effects. We perform and average two cycles of power- and temperature-dependent STEM-EELS to control current and hysteresis effects, reported in Figs. S10 and S11 and Table S1 and S2.

We first analyze our photomodulated low-loss measurements by calculating image-averaged differential spectra, which facilitates interpretation of the subtle spectral changes (Fig. 4B). The pronounced amplitude modulation over 2–8 eV likely reflects a near-surface electromagnetic (EM) response associated with surface polaritonic excitations, rather than only a static field or plasma. Denoted as "EM field," this signal may arise from surface phonon polaritons or surface plasmon-phonon polaritons (SPPs) commonly reported in $SrTiO_3$ (*51–53*). Its intensity depends on both the electron-beam current and the laser irradiance, consistent with beam coupling to photoinduced near-surface modes. These excitations extended 1–3 pixels beyond the dark-field image intensity and the spatially-resolved carrier and heating signatures. Prior Lorentz imaging measurements on $SrTiO_3$:Rh mapped phase shifts associated with photoinduced fields with high precision (*54*), whereas the present STEM-EELS measurements additionally probe electron energy loss into these excitations and therefore provide more direct information about the character of the states contributing to the near-surface response.

Temperature-dependent effects manifest in both the photomodulated and heated STEM-EELS differential spectra (Fig. 4B). The differential peaks from the heating control can be directly mapped onto the photomodulated spectrum, and nearly all the differential amplitude corresponds to thermal effects. These thermal spectral shifts are a result of the $SrTiO_3$ thermal lattice expansion, used as a contrast mechanism in prior EELS thermometry studies (*9*–*11*). We employ temperature-dependent TDDFT calculations to investigate the spectral shifts induced by thermal lattice expansion (Fig. S12). The multi-mode $SrTiO_3$ peak shifts are more complex than that of single-element metals and semiconductors reported in prior plasmon thermometry studies (*9*, *10*).

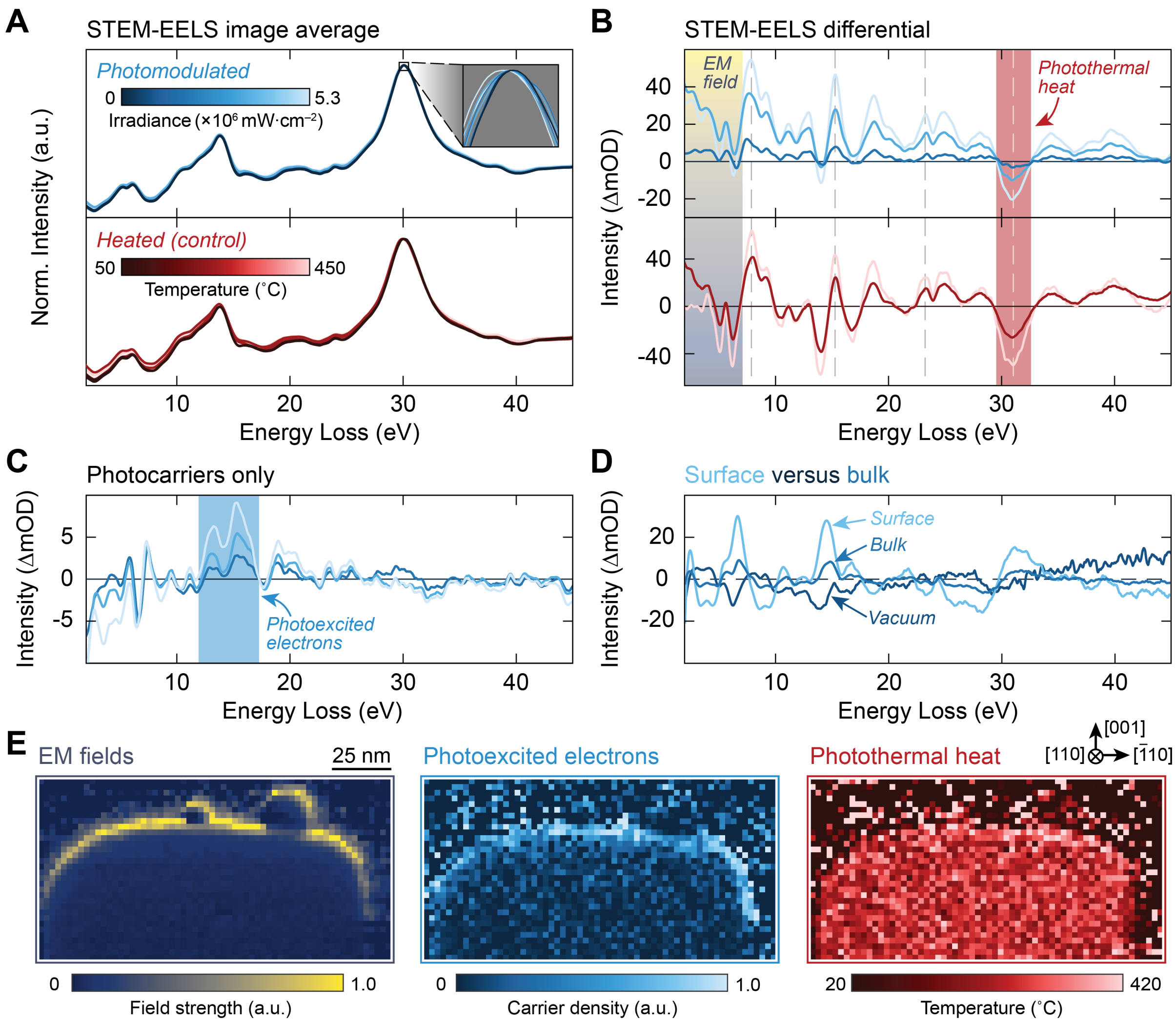

**Fig. 4. Photomodulated STEM-EELS imaging.** (**A**) Laser power- and temperature-dependent STEM-EEL spectra averaged across a spectral image of the $SrTiO_3$:Rh/Cu nanoparticle in Fig. 2. The color bars indicate the irradiance and temperature set throughout two hysteresis cycles, and the inset highlights the redshift of peak 2. (**B**) Corresponding differential spectra averaged over both cycles. Differential spectra were calculated using reference "ground state" spectra at 0 $mW\cdot cm^{-2}$ irradiance and 50 ˚C temperature, respectively. Spectral regions used to map local EM fields and photothermal heating are highlighted. (**C**) Photomodulated EEL spectra for $5.3\times10^6$ $mW\cdot cm^{-2}$ laser irradiance with EM field and photothermal heat backgrounds subtracted. The spectral region used to map photoexcited electrons is highlighted. (**D**) Differential photomodulated spectra at select regions of the photomodulated, background-subtracted STEM-EELS image. Arrows indicate the region on the nanoparticle where spectra were acquired. (**E**) Photomodulated STEM-EELS maps for $5.3\times10^6$ $mW\cdot cm^{-2}$ laser irradiance. Angstrom-scale EM fields, photoexcited electrons, and photothermal heat are mapped using the differential spectral amplitude over spectral regions highlighted in (B) and (C).

The effects of photocarriers become more apparent after subtracting the EM-field and photothermal-heating backgrounds (Fig. 4C). The primary feature in the differential spectrum is a positive signal at peak 1 induced by photocarriers. The differential signal induced by the EM field (<8 eV) and the photocarriers (12–16 eV) matches well with our ab initio TDDFT calculations of photocarriers in $SrTiO_3$ (Fig. S13). We model the influence of photoexcited electrons and holes using two chemical potentials, one for electrons in CBs and one for holes in VBs, by using constrained density functional perturbation theory (cDFPT) as an input for our TDDFT calculations (see Materials and Methods) (*55*). EM-field and photothermal-heating background fits are shown in Fig S14.

The angstrom-scale STEM-EELS spectra vary across the nanoparticle due to heterogeneity of Cu cocatalysts, surface oxygen vacancies, and vacuum background (Figs. S15–S17). To isolate photoexcited carriers from these spatially varying spectral backgrounds, we implement a rigid-shift analysis at every pixel in the photomodulated STEM-EELS image. For each pixel, we begin with the background-subtracted photomodulated differential spectrum, calculated as the natural logarithm acquired with the laser on divided by laser off, where the laser power is set to $5.3\times10^6$ $mW\cdot cm^{-2}$. We then rigidly shift the underlying ground-state spectrum to best match the differential spectrum's spectral intensity. The best match is then directly subtracted from the photomodulated differential spectrum. As a result, we remove pixel-dependent heating effects resulting from the photothermal heating spectral redshift. Averaging spectra in the thermal and background-corrected photomodulated image shows that the photocarrier differential signal increases at the nanoparticle's surface, and the nanoparticle's bulk signal amplitude is comparable to the vacuum without photocarriers (Fig. 4D).

We spatially map the differential spectral amplitude at the highest $5.3\times10^6$ $mW\cdot cm^{-2}$ laser irradiance (Fig. 4E). The nanoparticle is imaged along the [110] zone axis with 14 Å STEM-EELS resolution. We resolve an EM field strength highest at the nanoparticle's [001] facet, like prior Lorentz imaging studies (Fig. 4E, left panel) (*54*). Photoexcited electrons are measured to be localized at the [001] facet where the SSPs EM field is highest (Fig. 4E, middle panel). We quantify the photocarrier density by integrating the spectral intensity predicted by cDFPT turboEELS calculations, reporting a surface photocarrier density of $\sim1\times10^{19}$ $cm^{-3}$ at this irradiance, ~70% greater than the bulk photocarrier density. Fig. S18 compares the relative intensity of the measured dark-field signal, differential EM fields, temperature, and photocarrier density as quantified by the STEM-EELS mapping along the [001] and [$\bar{1}$10] zone axes. The relative contribution from electrons and holes is difficult to quantify due to the inseparability of electrons and holes in our cDFPT model. However, photoexcited electrons more dramatically influence the low-loss spectrum by introducing new modes after filling previously unoccupied CB states. The nanoparticle's uniform photothermal heating serves as an experimental control, which we map using the raw differential intensity of peak 2 in Fig. 4B. Thermal dissipation reaches an equilibrium at nanometer length scales in the steady state, so any signal fluctuations within the nanoparticle result from signal-to-noise limits (Fig. 4E, right panel). Additional experimental controls for the spatially resolved SP, image drift, nanoparticle thickness, and the energetic shifts of peak 2 are presented in Figs. S19–22. Supplementary Movies S2–S6 highlight these irradiance-dependent differential shifts.

That is not to say photomodulated STEM-EELS is without challenges. First, access to an optically coupled STEM with high spatial and energy resolution remains a significant barrier. Although recent advances in direct electron detectors and imaging filters greatly improve signal-to-noise, long acquisition times are still required. These acquisitions also entail precise specimen spatial drift control and minimal ZLP energy drift (see Supplementary Text). Second, interpreting differential STEM-EEL spectra is analytically challenging, aided here by TDDFT calculations. It is already difficult to distinguish bulk plasmons from interband transitions in the loss function. Our TDDFT calculations of $SrTiO_3$'s dielectric function emphasize the complex interplay between interband transitions and bulk plasmons. Although the real component of the dielectric function does not cross zero, lossy and screened plasmon modes likely exist considering the absence of high-energy absorption features (*56*). Third, the specimen must remain stable under electron irradiation, resistive heating, and photoexcitation. In this work, the beam current and scan rate were not observed to form oxygen vacancies during irradiance-dependent hysteresis cycles, and pretreatment conditions with high power laser exposure and 450 ˚C heating were conducted in vacuum to further reduce structural changes during STEM-EELS imaging (*57*).

**Discussion**

The development of photomodulated STEM-EELS introduces new spatial limits for imaging photoexcited carrier and thermal distributions. The combined spectral and spatial resolutions exceed those of existing imaging techniques, including scanning probe microscopy, optical interferometric scattering, Lorentz microscopy, and EELS thermometry. Our TDDFT calculations project $SrTiO_3$'s EEL spectrum onto its band structure. This new theoretical advance is critical for a deeper interpretation of the convoluted features in low-loss spectra. Here, we utilize these methods to quantitatively map the photoexcited electron density in $SrTiO_3$:Rh/Cu—directly imaging the photocarrier-induced EM fields extending beyond the nanoparticle's surface, as well as photothermal heating uniform across the nanoparticle's bulk. Our photomodulated STEM-EELS imaging results agree well with the reported mechanisms of preferential photocarrier transport along $SrTiO_3$'s [001] facet and trapping at oxygen vacancy sites (*3*, *24*). However, our results indicate that significantly fewer photocarriers are transferred into the Cu cocatalysts and trapped at Rh dopants than previously theorized. Our results suggest that unintended surface trapping contributes to lower-than-expected photocatalytic performance, indicating a need for synthetic design rules that suppress surface-trap formation while improving carrier accumulation at co-catalysts to enhance solar $H_2$ evolution. This photomodulated STEM-EELS approach offers a general framework to image carrier- and thermal-induced phenomena in nanoscale structures.

**Acknowledgments:**

We thank Dr. Alex Ye for alternative TEM grid preparation. The computations presented here were conducted in the Resnick High Performance Computing Center, a Resnick Sustainability Institute facility at the California Institute of Technology.

**Funding:**

This research was supported as part of the Ensembles of Photosynthetic Nanoreactors, an Energy Frontier Research Center funded by the U.S. Department of Energy, Office of Science under Award No. DE-SC0023431. Work performed at the Center for Nanoscale Materials, a U.S. Department of Energy Office of Science User Facility, was supported by the U.S. DOE, Office of Basic Energy Sciences, under Contract No. DE-AC02-06CH11357. Sandia National Laboratories is a multi-mission laboratory managed and operated by National Technology and Engineering Solutions of Sandia, LLC., a wholly owned subsidiary of Honeywell International, Inc., for the U.S. Department of Energy's National Nuclear Security Administration under contract DE-NA0003525. L.D.P. was supported by the National Science Foundation Graduate Research Fellowship under Grant No. DGE-1745301 and the Office of Science Graduate Student Research (SCGSR) program. The SCGSR program is administered by the Oak Ridge Institute for Science and Education for the U.S. Department of Energy, Office of Science under contract No. DE-SC0014664. W.L. acknowledges support from the Korea Foundation for Advanced Studies. The views expressed in this article do not necessarily represent the views of the U.S. Department of Energy, National Science Foundation, or the United States Government.

**Author contributions:**

Conceptualization: LDP, WL, TEG, SKC

Methodology: LDP, WL, TEG, SKC

Software: LDP, WL, SKC

Validation: LDP, WL, PRP, HHW, TEG, SKC

Formal analysis: LDP, HHW

Resources: LDP, WL, BWL, ZC, JW, YL, AK, TEG

Investigation: LDP, WL, JW, YL, TEG

Visualization: LDP, WL, PRP, HHW, TEG

Funding acquisition: SA, SKC

Project administration: SA, TEG, SKC

Supervision: XP, AAT, SA, AK, JPP, TEG, SKC

Writing – original draft: LDP, WL, PRP, HHW, SA, TEG, SKC

Writing – review & editing: LDP, WL, PRP, HHW, BWL, ZC, JW, YL, HHW, XP, AAT, AK, SA, JPP, TEG, SKC

**Competing interests:** The authors declare that they have no competing interests.

**Supplementary Materials**

Materials and Methods

Supplementary Text

Figs. S1 to S23

Tables S1 to S2

Movies S1 to S6

# Supplementary Materials for

## Imaging nanoscale photocarrier traps in solar water-splitting catalysts

Levi D. Palmer,[1] Wonseok Lee,[1] Pushp Raj Prasad,[2] Bradley W. Layne,[2] Han-Hsuan Wu,[3] Zejie Chen,[2] Jianguo Wen,[4] Yuzi Liu,[4] Xiaoqing Pan,[3,5,6] A. Alec Talin,[7] Akihiko Kudo,[8] Shane Ardo,[2,6,9] Joseph P. Patterson,[2,6] Thomas E. Gage,[4]* Scott K. Cushing[1]*

Corresponding author: scushing@caltech.edu (S.K.C.) & tgage@anl.gov (T.E.G.)

**The PDF file includes:**

Materials and Methods

Supplementary Text

Figs. S1 to S23

Tables S1 to S2

**Other Supplementary Materials for this manuscript include the following:**

Movies S1, S2, S3, S4, S5, S6

## Materials and Methods

Nanoparticle synthesis and preparation for TEM imaging

Rhodium-doped (1 wt% doping) strontium titanate nanoparticles were synthesized according to previously reported solid-state reaction syntheses (*23*). To prepare particles for TEM imaging, a 0.1 mg/mL nanoparticle suspension solution was created in filtered, HPLC-grade ethanol. The solution was vortexed and then sonicated for 60 min to break up large aggregates. The solution was then vortexed again and left to sit for 5 minutes such that large aggregates condense at the bottom of the aliquot tube. Then, the supernatant was removed and centrifuged to condense non-aggregated nanoparticles. This new supernatant was removed and discarded, leaving an infranatant with condensed nanoparticles. This infranatant was mixed using a pipette. A 5 µL drop of this infranatant solution was then drop-cast onto a high-mesh Cu TEM grid with a 3–5-layer graphene support. All prepared nanoparticles on TEM grids were heated in vacuum at 450 ˚C for at least 30 minutes prior to imaging to mitigate carbon contamination from residual synthetic precursors.

Ground-state STEM-EELS and imaging

Ground-state STEM characterization was performed using a Thermo Fisher Scientific Spectra 300 microscope with a cold field-emission gun (ZLP fwhm ~0.43 eV) at 80 kV. For STEM-EELS, a 320 kx magnification, 38 mm camera length, and 8 mrad convergence semi-angle were utilized. A Gatan Continuum ER spectrometer with a CMOS detector (50 meV/ch dispersion) with a 2.5 nm or a 5 nm entrance aperture was used to acquire low-loss and core-loss measurements, respectively. Atomic-resolution integrated differential phase contrast (iDPC) imaging was performed with the Thermo Fisher automated iDPC analysis in Velox.

Photomodulated STEM-EELS

Photomodulated STEM-EELS measurements were performed on a JEOL NEOARM microscope with a cold field-emission gun (ZLP fwhm ~0.5 eV). STEM-EELS data was acquired at 200 kV with 500 kx STEM magnification mode. Measurements were performed at ∼0° tilt, approximately along the [110] zone axis. The beam current was approximately 170 pA or 16,000 pA/nm$^2$ in STEM. The Gatan Continuum Imaging Filter was aligned using a 2.5 mm entrance aperture with 30 meV/ch dispersion and a Gatan K3 direct electron detector. The Gatan subscan feature (32 × 32 points/px) was enabled to diffuse the beam over the scanning region on the specimen, reducing beam-induced heating and damage. The acquisition time was fixed at a 0.25 s view time in dualEELS (0.7% live time low-loss with 0 eV shift, 100% live time high-loss with 7 eV shift). Photomodulation was performed using the IDES Cobolt C-FLEX (C4) system from Hubner Photonics. The 405 nm (Cobolt 06-MLD) laser is integrated into the scope using two mirrors after the fiber, is focused using a UV-fused silica lens, passes through a UV-fused silica window, and is directed to the specimen using a mirror in the microscope (IDES Luminary Micro). Approximately 35% of the laser power reaches the specimen after losses. The laser power was swept from 300 mW to 0 mW and back with 100 mW steps according to Table S1.

A Gatan 652 TA bulk furnace heating holder with ±5 ˚C accuracy of the controller setpoint was used for temperature-dependent STEM-EELS. The specimen was first loaded into a vacuum pumping station and heated to 450° C for 30 min to minimize surface carbon contamination and again heated to 450° C for 30 min inside the microscope column to minimize transfer-induced contamination. During STEM-EELS measurements, the holder temperature was first set to 450° C and the temperature was cycled down in 200° C steps to 50° C and then thermally cycled back up to investigate heat-induced hysteresis.

Core-loss EELS theory

To simulate core-loss $SrTiO_3$ Ti $L_{2,3}$ edge and O K edge, as well as the Cu and $Cu_2O$ Cu $L_{2,3}$ edge, we leverage the obtaining core excitations from the ab initio electronic structure and the NIST Bethe–Salpeter equation (BSE) solver (OCEAN) package. OCEAN utilizes Quantum ESPRESSO as part of its workflow to perform density functional theory (DFT) calculations of each material's electronic structure. OCEAN subsequently calculates the screening, core-hole exciton effects, and core-level spectra through BSE calculations. OCEAN 3.0.1 and Quantum ESPRESSO 7.4 were used for all calculations (*29*, *44*).

Typical DFT parameters include a 250 Ry wave function cutoff energy and $1.1 \times 10^{-9}$ Ry convergence threshold following structural relaxation and convergence testing. The atomic positions were variable-cell relaxed using $SrTiO_3$, Cu, and $Cu_2O$ crystallographic data. Self-consistent field calculations utilized a $6 \times 6 \times 6$ k-point mesh or greater, and non-self-consistent field calculations utilized an $8 \times 8 \times 8$ k-point mesh or greater. Pseudopotential files were produced using the ONCVPSP code and calculated using a PBE functional and nonlinear core correction (*30*). Typical BSE parameters include a $6 \times 6 \times 6$ x-mesh or greater. Input files can be found online (*58*).

Low-loss EELS theory

A 250 Ry cutoff energy and a $1 \times 10^{-12}$ Ry convergence threshold were utilized for self-consistent field calculations with a $14 \times 14 \times 14$ k-point mesh, 50 electronic states (bands), and standard atomic positions from crystallographic data. Pseudopotential files were again generated using the ONCVPSP code. The turboEELS solver requires ultrasoft or norm-conserving pseudopotentials. turboEELS calculates the low-loss volume plasmon spectra using either a Liouville–Lanczos or Sternheimer approach to linearized TDDFT. By calculating the charge-density susceptibility with the quantum Liouville equation, the inverse of the imaginary/longitudinal component of the dielectric function (the EEL spectrum) is simulated. Fixed momenta must be simulated to avoid local minima, included as $k_x = 0.085$ Å$^{-1}$ and $k_y$, $k_z = 0$ up to 2000 iterations for the linear response control. The linear response input of 2000 exact coefficients are read from the file up to 20000 iterations, extrapolated with an oscillatory function and 0.05 eV broadening. The Liouville–Lanczos solver is used unless specified. Input files can be found online (*58*).

To simulate photoexcited electrons and holes in $SrTiO_3$ and their influence on the low-loss spectrum, we employ constrained density-functional perturbation theory (*55*). Their influence on the low-loss EEL spectrum was modeled by varying the density of photoexcited carriers in the respective conduction and valence bands under the assumption that photoexcited carriers in the steady state are fully thermalized. Self-consistent field parameters were adjusted to balance computational expense as follows: 250 Ry cutoff energy, 35 electronic states, a $1 \times 10^{-12}$ Ry convergence threshold, $4 \times 4 \times 4$ k-point mesh. Smeared occupations are required, set to 0.0025 Ry degauss and a 0.01 Ry conduction band degauss. The number of electrons/holes (i.e., the chemical potential) was variable from 0 to 0.002 electrons per atom. The structural parameters were relaxed between simulated chemical potentials and no change in parameters was observed.

Photothermal effects, by contrast, were incorporated through lattice heating. Assuming isotropic lattice expansion, the low-loss EEL spectra at different lattice temperatures were calculated using lattice parameters determined at each temperature from the thermal expansion coefficient of $SrTiO_3$ in the turboEELS calculations.

Our low-loss calculations primarily employ the Liouville–Lanczos approach to calculate ground- and excited-state low-loss EEL spectra of $SrTiO_3$ because of its computational efficiency. However, the Sternheimer equation, which describes the linear response of the electron density to an external perturbation, was implemented to analyze the ground-state electron distribution in reciprocal space, described as the spectral projection in Fig. 3 in the main text. This formalism enables the direct computation of the perturbed wavefunction by solving a system of linear equations that incorporates the external perturbation, which is not feasible within the Liouville–Lanczos approach (*49*, *50*). From the perturbed wavefunction, both the perturbed electron density and its reciprocal distribution are obtained. By examining them, we identify the specific core and valence electronic states that contribute to the energy-loss processes, including plasmon excitations and interband transitions. Our calculations reveal key details about which electronic transitions and modes yield the low-loss EEL spectrum.

## Supplementary Text

Zero-loss peak alignment

The zero-loss peak was aligned using a center-of-mass fit over a 150 pixel window centered at the highest amplitude of the low-loss spectrum. The low-loss and high-loss spectra are both corrected for the ZLP energy for each pixel with a 7 eV spectral shift applied to the high-loss spectrum according to the set drift tube voltage for dualEELS.

Atomic STEM iDPC image analysis

Atomic resolution iDPC images were acquired using a Thermo Fisher Panther segmented detector. The nanoparticle is aligned along the [110] zone axis using a double-tilt holder and the auto-tilt Velox feature when resolving the Kikuchi pattern. However, post-collection analysis of the iDPC image illustrates subtle off-zone axis tilt due to misalignment of the measured lattice parameters and atomic spacing compared to bulk $SrTiO_3$. As mentioned in the main text, we map the Sr and Ti atomic columns by identifying circular atomic features via template matching and fit their position with a 2D Gaussian profile. We analyze the fit positions to calculate the interatomic distances between Ti–Ti / Sr–Sr and Sr–Ti (main text Fig. 2E). The measured interatomic distances are corrected uniformly such that measured interatomic distances in the $SrTiO_3$ nanoparticle bulk match the reported crystallographic bond lengths. This is performed by rescaling the crystallographic distances such that the ideal long bond (Ti–Ti / Sr–Sr) matches the observed long bond in the bulk by $\cos(\Delta\theta)$, a factor equivalent to the averaged measured distance divided by the crystallographic bond distance. The entire map of fit atomic positions, along with the analyzed region and the acquired iDPC image, are shown in Fig. S5.

Automatic Gatan spatial drift correction

Specimen drift was corrected using Gatan's automated live spatial drift correction. This built-in feature corrects for drift by adjusting the beam deflectors based on a cross-correlation image over a reference region. The cross-correlation quantifies the drift and an offset to the beam scan coordinates is applied. For the photomodulated and temperature-dependent STEM-EELS measurements in this study, drift is corrected once every row of the image, or every 62 pixels. Without drift correction enabled, the photomodulated STEM scan would have otherwise moved ~6 nm. This drift relates to ~0.2 nm drift per row or ~0.1% of each pixel on average between drift-corrected measurements.

Laser alignment procedure

Laser alignment onto the electron beam's optical axis requires a multi-stage process. First, IDES performs initial alignment of the laser and optics with the tool at atmosphere, which is typically performed during the tool's initial installation. Next, sub-millimeter alignment precision is achieved using the IDES optical alignment holder within the STEM side-entry port with the microscope under vacuum. Next, a series of low-to-high magnification alignment stages are performed to progressively increase the alignment precision to ~100 micrometer accuracy. This is achieved by using the laser to burn a hole in an amorphous carbon foil. Finally, the beam shift is corrected between low- and high-magnification imaging modes to ensure that the electron beam does not drift away from the laser spot aligned on the specimen.

Beam current stability and background subtraction

The beam current of the microscope is variable due to oxidation of the cold field-emission gun, influenced by its vacuum quality. To refresh the current, either a high-temperature or low-temperature flash is performed. The tip was typically low-T flashed prior to the laser power or temperature hysteresis cycles. The estimated beam current was ~170 pA at the specimen. Additionally, the variable background induced by spatial and beam-current-dependent fluctuations of the zero-loss peak was removed using a power-law fit. The subtracted fit backgrounds of the image-averaged data are shown in Fig. S14.

Irradiance (power density) calculations

The laser irradiance is calibrated according to the set laser power and appreciable losses due to the fiber-coupled housing and optics used for coupling into the STEM.

Photomodulated STEM-EELS analysis and rigid shift model

All STEM-EELS data was analyzed using the HyperSpy Python loader; analysis scripts and data are available online (*58*).

Spatially resolved maps were generated from energy-aligned low-loss EELS spectrum images, where the ZLP in each pixel of each spectrum image was fit using center-of-mass analysis and aligned to 0 eV. Spectrum images were acquired under pump-off and pump-on conditions. The ZLP-aligned datasets have dimensions corresponding to pump condition, spatial row, spatial column, and energy loss. For each spatial pixel, spectra from repeated pump-off measurements were averaged using the indices corresponding to 0 mW excitation, while spectra from repeated pump-on measurements were averaged using the indices corresponding to 300 mW excitation (results for all pump-on conditions are shown in the supporting movies).

At each pixel, the pump-off and pump-on spectra were first normalized to the maximum spectral intensity within the 13–17 eV energy-loss window, which corresponds to the primary plasmon for photocarrier excitations. This normalization was applied to reduce the influence of pixel-to-pixel intensity variations of the probe's descan alignment and to correct for differences in total signal level between spectra. A logarithmic differential spectrum was then calculated as:

$$\Delta(E) = ln\left(\frac{I_{300}(E) + \epsilon}{I_0(E) + \epsilon}\right) \times 1000$$

where $I_{300}(E)$ and $I_0(E)$ are the normalized pump-on and pump-off spectra, respectively, and $\epsilon = 10^{-10}$ was added to avoid division by zero. The differential spectrum was smoothed using a Savitzky–Golay filter with a 61-point window and third-order polynomial. The spectrum was then baseline-offset by subtracting the value of the differential spectrum at the largest plasmon peak's tail at approximately 33 eV, forcing the tail of its differential response to be aligned to zero at 33 eV. The same normalized differential spectrum was used to generate the initial EM-field and heating maps, but the maps were obtained by integrating over different energy-loss windows.

EM-field map. The EM-field map represents the spatial distribution of the pump-induced low-energy response outside the nanoparticle and is interpreted here as the EM-field-related response. This map was generated by integrating the smoothed pump-on minus pump-off differential spectrum over the 11–15 eV energy-loss range:

$$M_{\mathrm{EM}(i,j)} = \sum_{E=11-15\mathrm{eV}} \Delta(E, i, j)$$

where $M_{\mathrm{EM}(i,j)}$ is the integrated EM-field response at spatial pixel (*i,j*). This integration window captures the low-loss spectral region where the pump-induced optical/electromagnetic response is most apparent in the initial differential spectrum.

Heating map. A separate heating map was generated from the higher-energy region of the same differential low-loss spectra. For each pixel, the differential signal was integrated over the 31.5–33 eV energy-loss window:

$$M_{\mathrm{heat}(i,j)} = \sum_{E=31.5-33\mathrm{eV}} \Delta(E, i, j)$$

This higher-energy integration window was selected to track spectral changes associated with pump-induced heating corresponding to the most intense, higher-energy plasmon peak. After the heating map was generated, outlier pixels beyond two standard deviations of the image average were set to zero amplitude on the map. Nonzero values in the temperature map were used to calculate the mean and standard deviation of the nanoparticle's differential temperature. The differential was calibrated using the differential amplitude in the 31.5–33 eV energy-loss region from the resistive heating control experiments.

EM- and heat-corrected carrier map. Analysis was then performed to isolate the carrier-related spectral response after accounting for (1) beam-induced field and (2) thermal effects. This correction was necessary for two respective reasons. (1) The electron beam current induced visible fluctuations of the EM field background signal stemming as tails off the ZLP. (2) Pump-induced heating also shifted the low-loss peak at 13–17 eV. If not removed, these amplitude- and shift-induced differential features overlapped and obscured differential plasmon features used to quantify carrier-related signal.

For each pixel, the pump-off spectrum was again obtained by averaging the 0 mW spectra, and the high-power spectrum was obtained by averaging the 300 mW spectra. Both spectra were again normalized to their maximum intensity in the 13–17 eV range, and the differential spectrum was calculated as above. Each raw differential spectrum was again offset so the largest plasmon's tail at 33 eV was zero. To account for EM-field contributions of the fluctuating ZLP tail, a power-law background of the form:

$$B(E) = aE^b$$

was then fit over the 2–45 eV range and subtracted from the raw differential spectrum. The fit power-law background-subtracted differential spectrum was smoothed using a Savitzky–Golay filter with a 41-point window and third-order polynomial.

To model the thermal contribution, the normalized 300 mW spectrum was shifted rigidly along the energy axis by integer pixel offsets ranging from −6 to +6 pixels. For each tested shift, the shifted high-power spectrum was compared to the unshifted pump-off spectrum by calculating a simulated shift-induced differential spectrum:

$$\Delta_{shift}(E, d) = ln\left(\frac{I_{300}^{norm}(E + d)}{I_0^{norm}(E)}\right)$$

where *d* is the tested integer-pixel energy shift to correct for thermal effects. Each simulated shift-induced differential spectrum was also offset to zero at 33 eV. The shift that best reproduced the measured differential spectrum was identified by minimizing the summed squared residual between the measured background-subtracted differential spectrum and the simulated shift-induced differential spectrum over the 4–33 eV fitting window:

$$\chi^2(d) = \sum_{E=4-33\text{eV}} \left[\Delta_{bg}(E) - \Delta_{shift}(E, d)\right]^2$$

where $\Delta_{bg}(E)$ is the measured differential spectrum after the first power-law background subtraction. This fitting procedure determines, independently for every pixel, the rigid spectral shift that most closely accounts for the observed pump-induced differential response over the broad low-loss energy range.

The best-fitting rigid-shift differential spectrum was then subtracted from the measured background-subtracted differential spectrum:

$$\Delta_{clean}(E) = \Delta_{bg}(E) - \Delta_{shift}(E, d_{best})$$

This subtraction removes the component of the pump-induced differential signal that can be explained by a thermal redshift of the spectrum. In other words, this subtraction accounts for the spectral redshifts from thermal expansion that are typically measured using plasmon energy expansion thermometry. Importantly, the correction is performed at each pixel independently, allowing the magnitude and direction of the thermal shift to vary spatially across the spectrum image. Ideally, the heating control STEM-EELS datasets could be used for this purpose, but the spatial drift created by the resistive heating holder was too significant.

After subtracting the best-fit shift contribution, a second power-law background subtraction was applied over the 2–45 eV range. This second background subtraction removes any residual slope or slowly varying baseline introduced by the rigid-shift correction. The resulting corrected spectrum, $\Delta_{corr}(E)$, represents the residual pump-induced response after removal of the broad thermal-shift contribution and background.

The final thermally corrected carrier map was generated by integrating the negative of this corrected residual differential spectrum over the 12–16 eV energy-loss window:

$$M_{\text{carrier}(i,j)} = \sum_{E=12-16\text{eV}} -\Delta_{corr}(E,i,j)$$

Because the 300 mW spectra were rigid-shifted, instead of the non-photomodulated 0 mW spectra, a negative sign was applied to the spectra so negative-going residual differential features in the 12–16 eV range appear as positive carrier-map intensity.

A

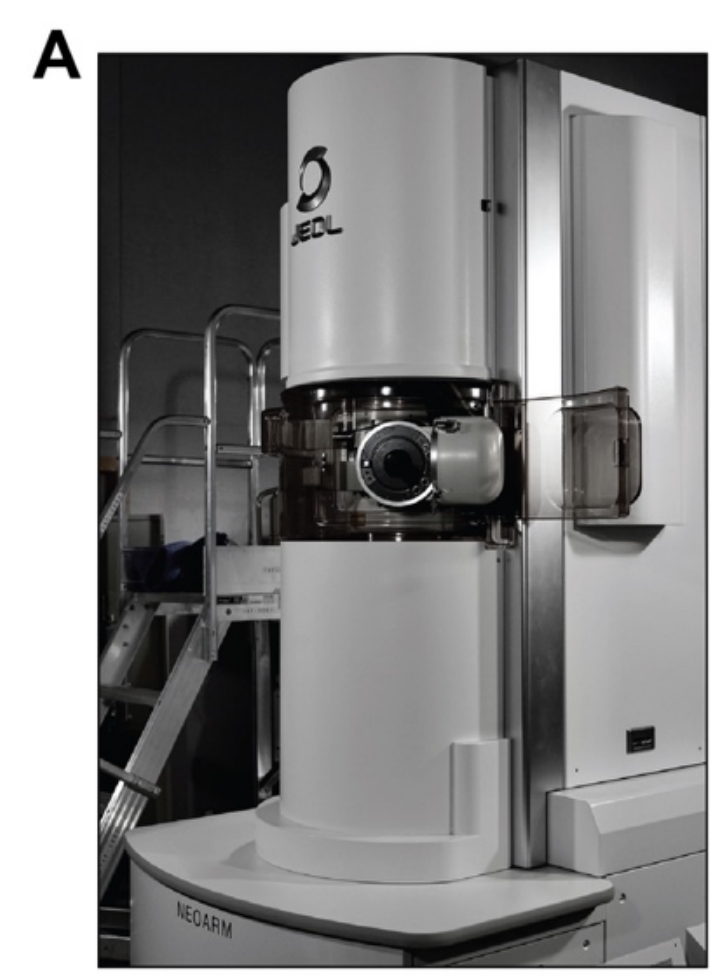

B

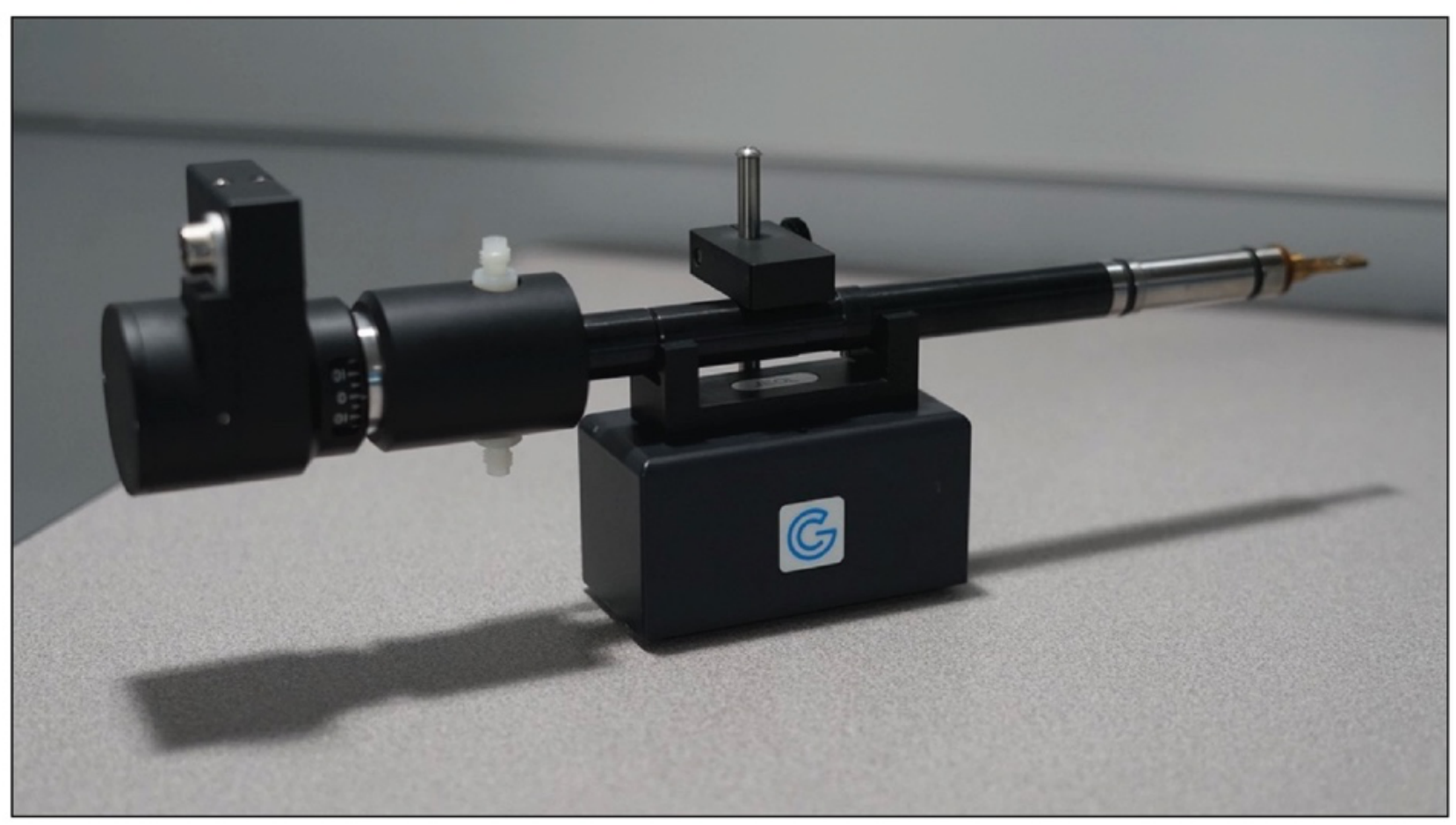

**Fig. S1.**

**Photomodulated STEM-EELS platform with in situ heating.** (**A**) The JEOL NEOARM microscope with optically coupled photoexcitation through the integrated IDES laser module. (**B**) The in situ bulk heating holder allows temperature-dependent control measurements through side-entry specimen insertion.

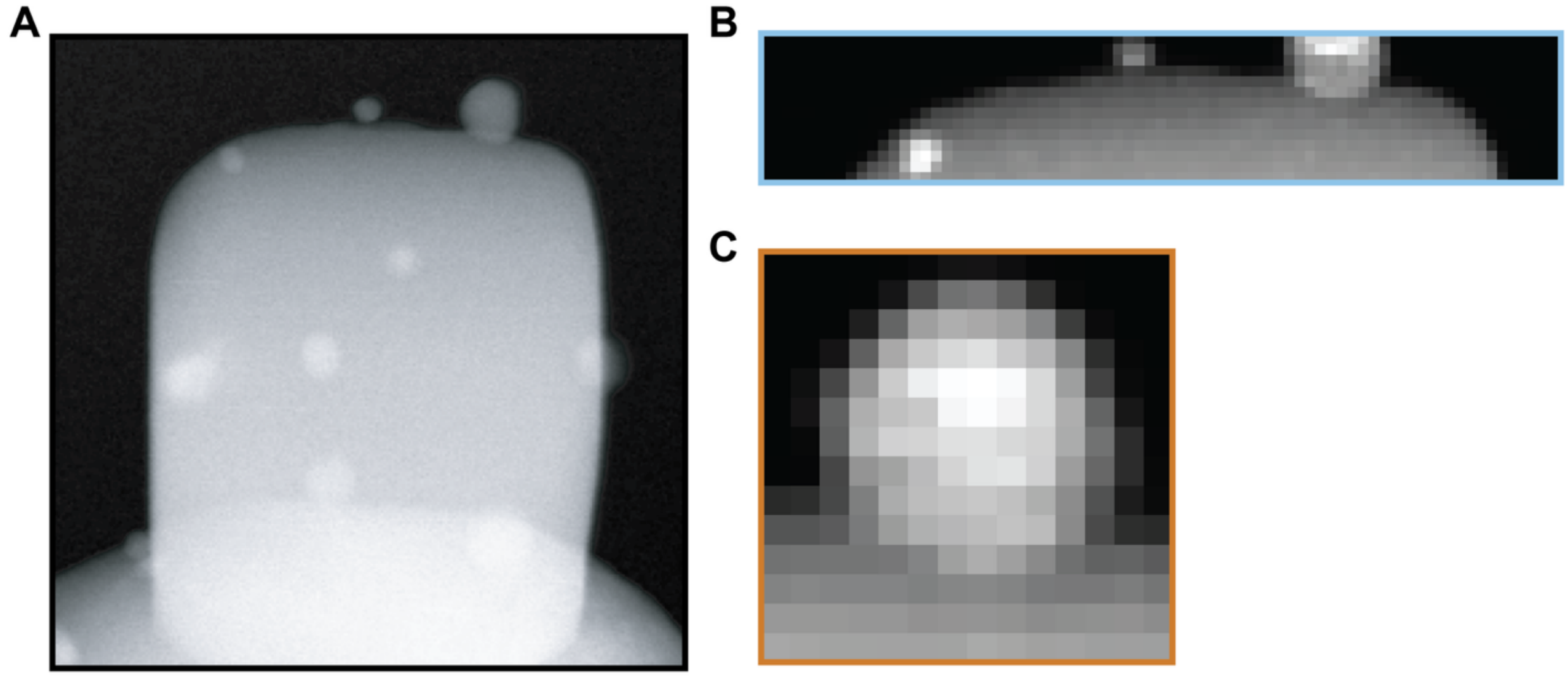

**Fig. S2.**

**ADF images of the Cu/$SrTiO_3$:Rh studied nanoparticle without false coloration.** (**A**) The $SrTiO_3$:Rh nanoparticle studied with low-loss EELS, sitting on a sintered aggregate chunk with distributed Cu nanoparticles. The vacuum region does not contain a support film. (**B**) The region of interest for the core-loss Ti $L_{2,3}$ and O K edge mapping. (**C**) The region of interest for the Cu $L_{2,3}$ edge mapping.

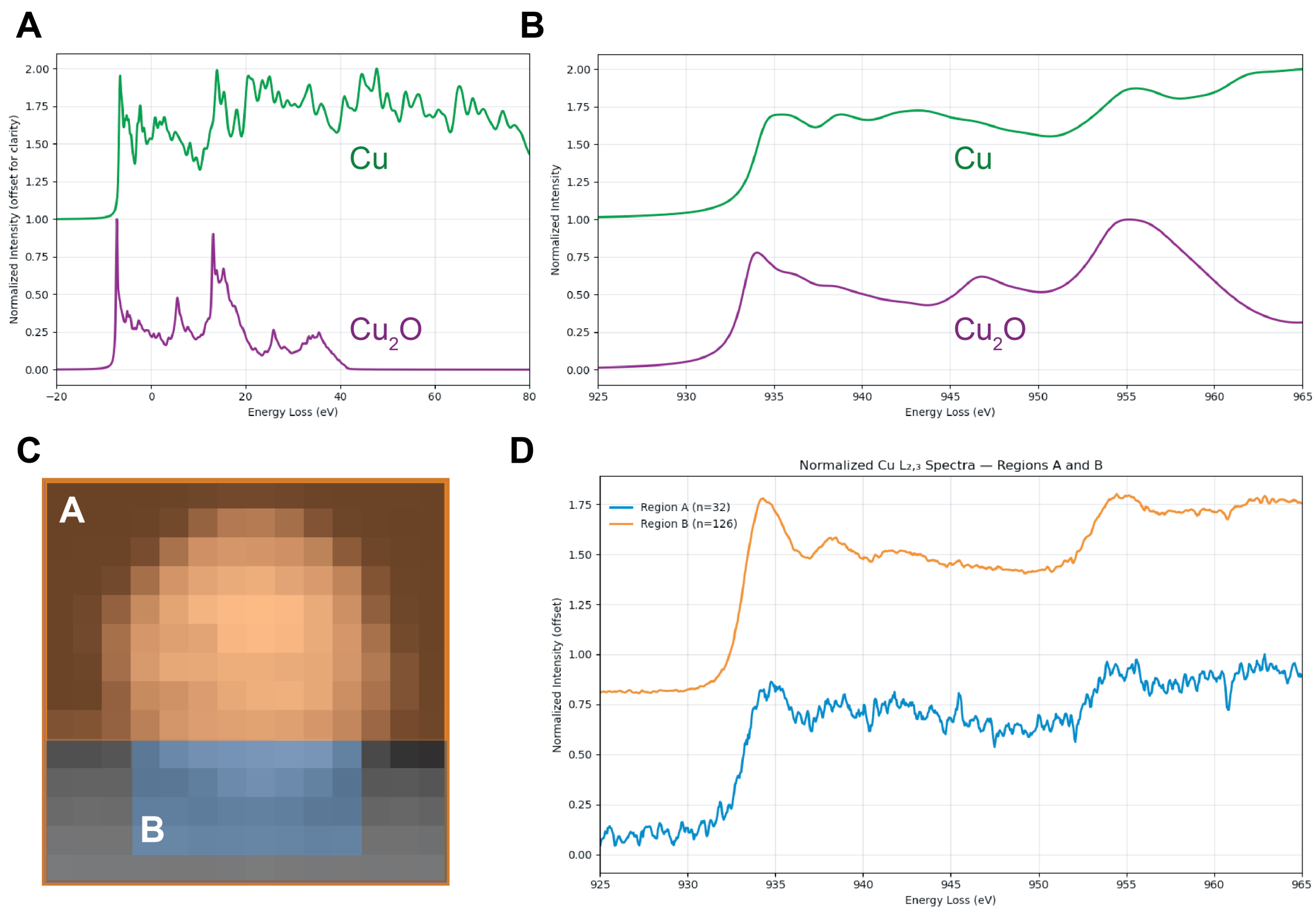

**Fig. S3.**

**Spatially resolved core-loss Cu $L_{2,3}$ edge spectra of the copper nanoparticle mapped in the main text.** (**A**) DFT+BSE calculated spectra for both Cu compositions. (**B**) The spectra as shown in (A) broadened according to lifetime broadening effects Cu(0) metal and $Cu_2O$. (C) The region of interest used for mapping, Region A and Region B are highlighted according to the corresponding colors in (**D**), which shows the image-averaged experimental STEM-EEL spectra.

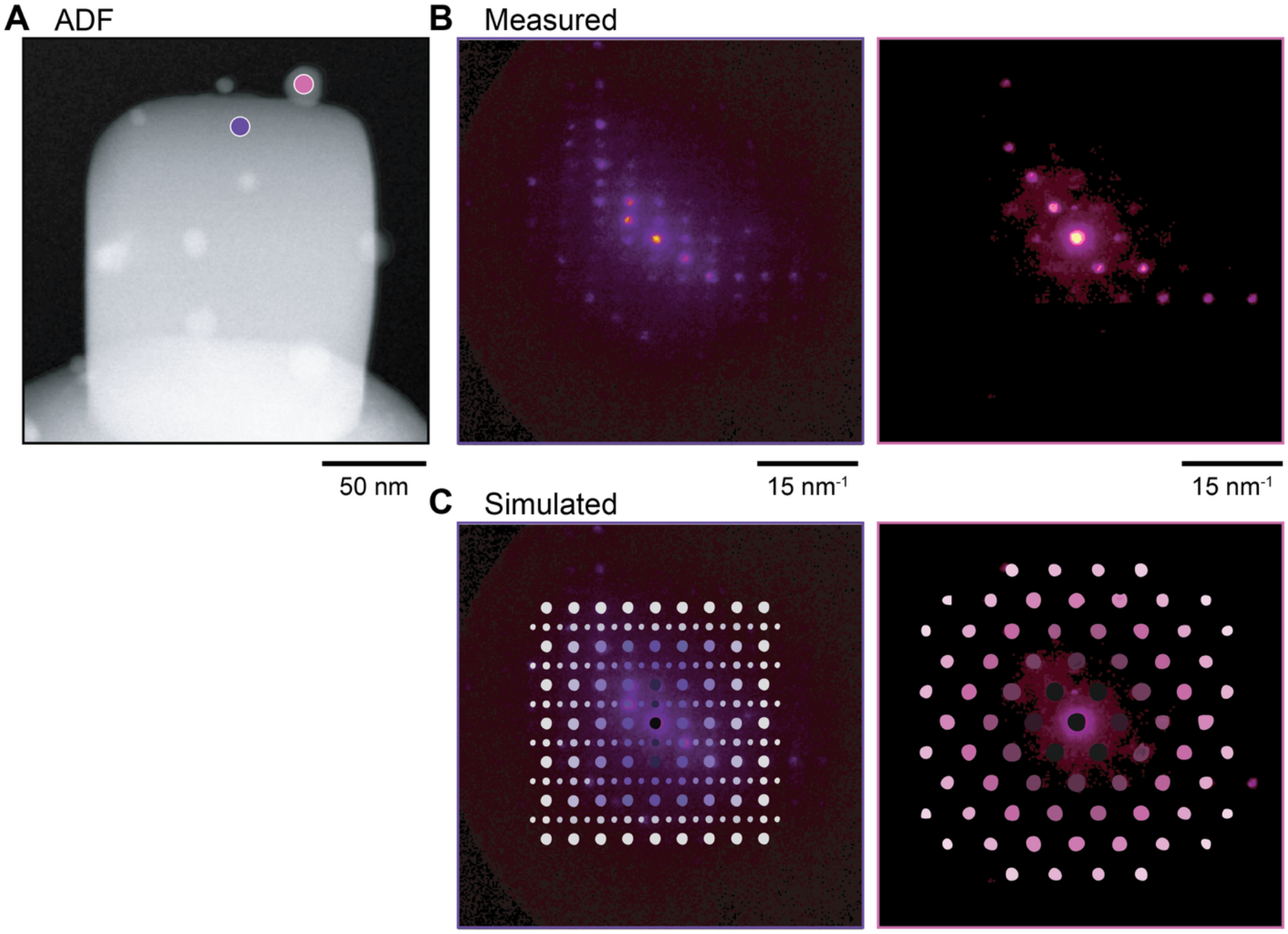

**Fig. S4.**

**Experimental and simulated Cu and $SrTiO_3$ diffraction patterns reflecting the [110] zone axis orientation.** (**A**) The region of interest with the nanoprobe diffraction spots indicated. (**B**) Measured nanoprobe diffraction patterns over $SrTiO_3$ (left in purple) and Cu (right in pink), optimized using the selected area electron diffraction aperture. Convergence angle of 1.2 mrad and an approximate probe size of 0.7 nm. (**C**) Corresponding simulated diffraction patterns according to the multislice method.

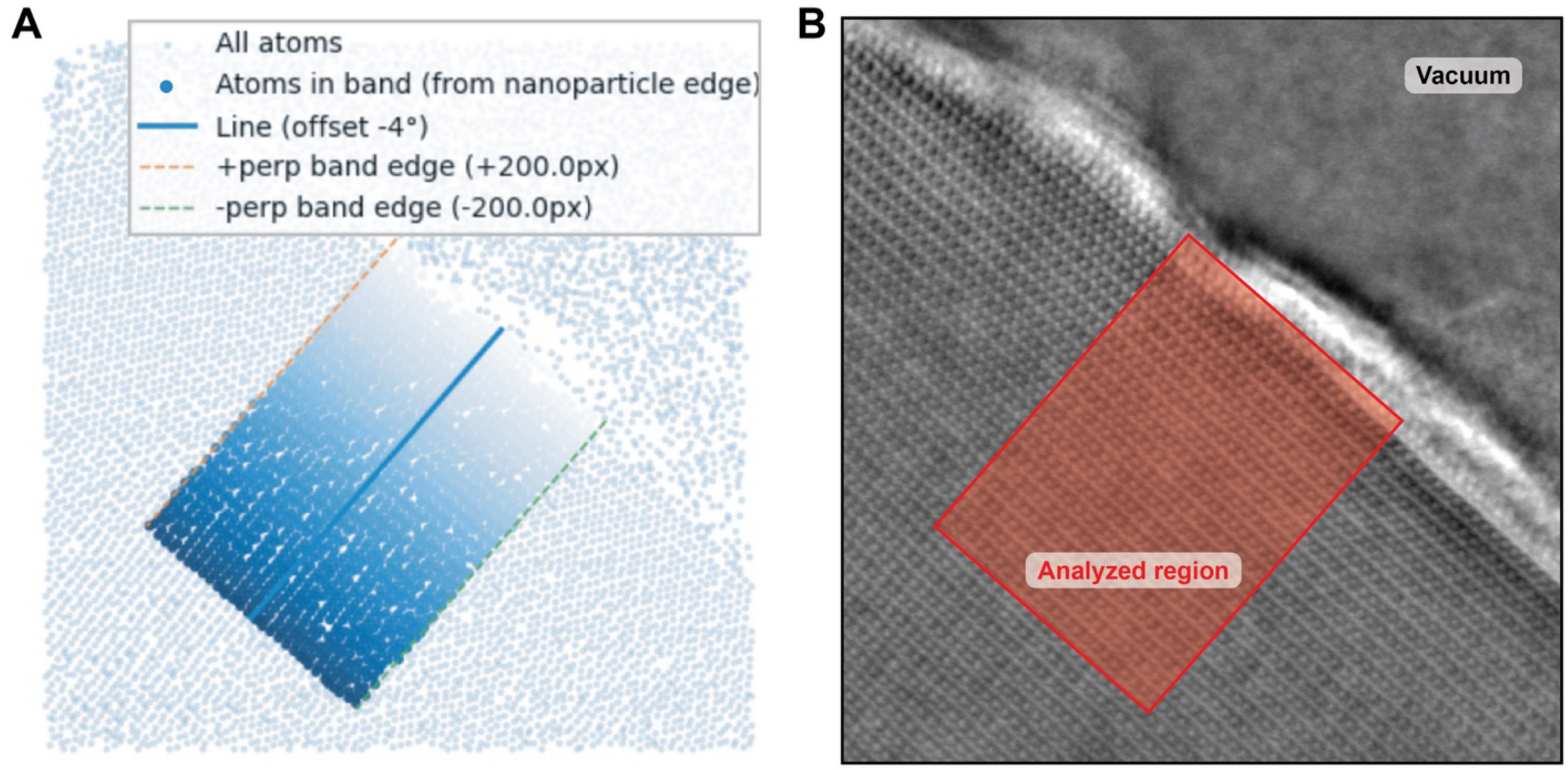

**Fig. S5.**

**Atomic resolution iDPC imaging and atomic position mapping.** (**A**) Gaussian fits of the atomic resolution iDPC map and the analyzed region within the image. This analyzed region is used for the statistical analysis in main text Fig. 2. (**B**) The iDPC image of as-synthesized, undoped $SrTiO_3$ nanoparticles. The analyzed and vacuum regions are highlighted.

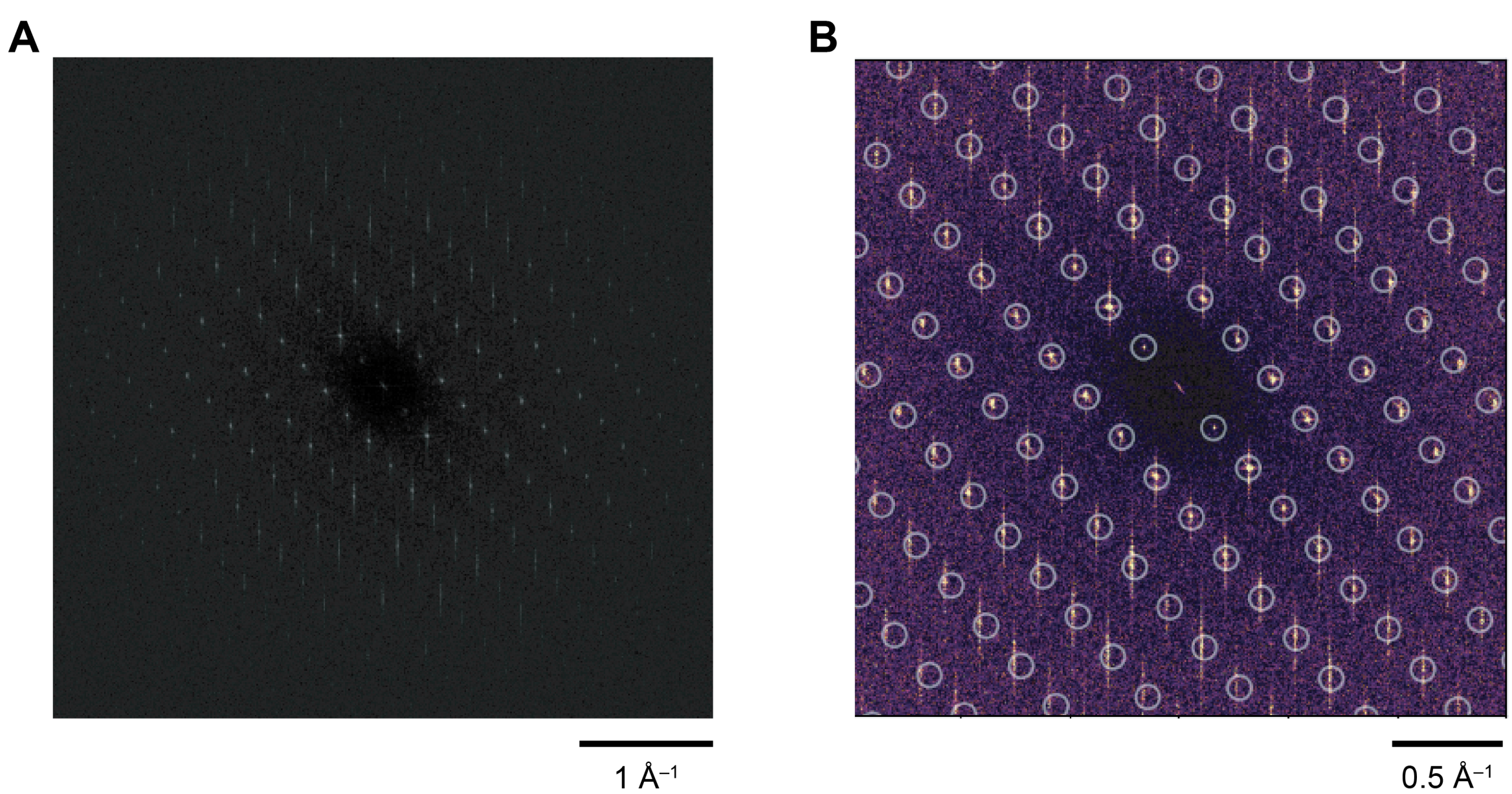

**Fig. S6.**

**Fourier transform of atomic position map.** (**A**) The virtual diffraction pattern of the atomic resolution imaging in Fig. S5B. (**B**) The virtual diffraction pattern color-scaled for visibility and the simulated [110] diffraction pattern of $SrTiO_3$ overlaid in white circles.

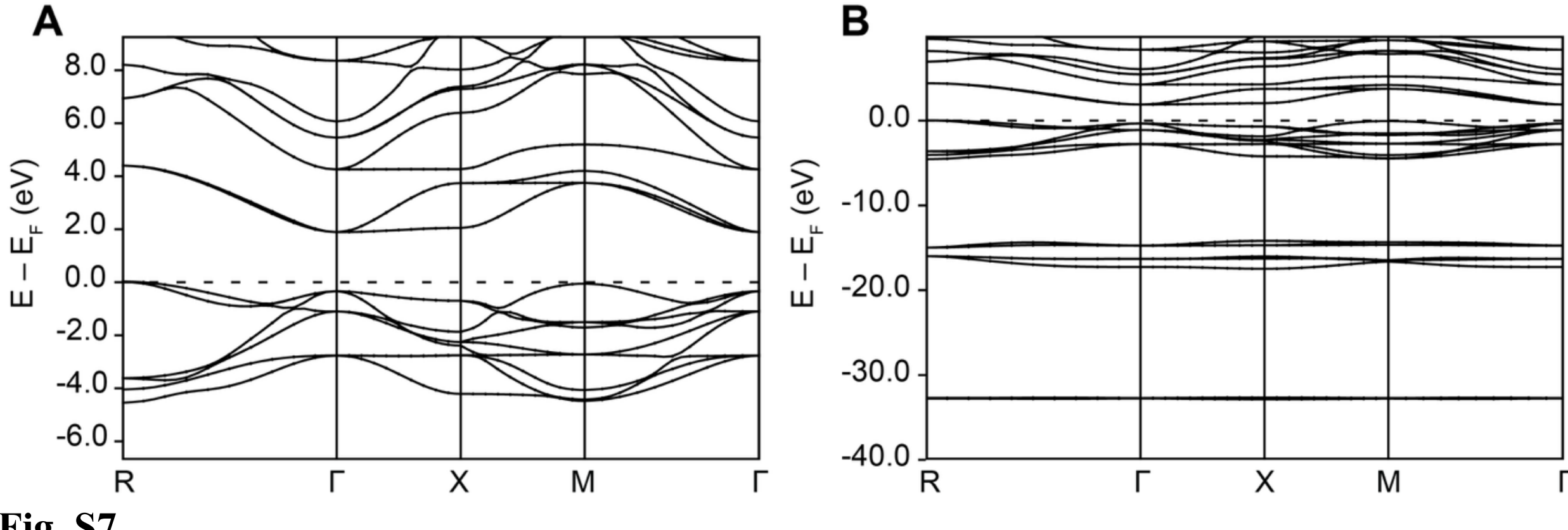

**Fig. S7.**

**$SrTiO_3$ band structure.** The DFT-calculated band structure of undoped, bulk strontium titanate to better visualize (**A**) the valence states and (**B**) the valence and core states as compared to main text Fig. 3.

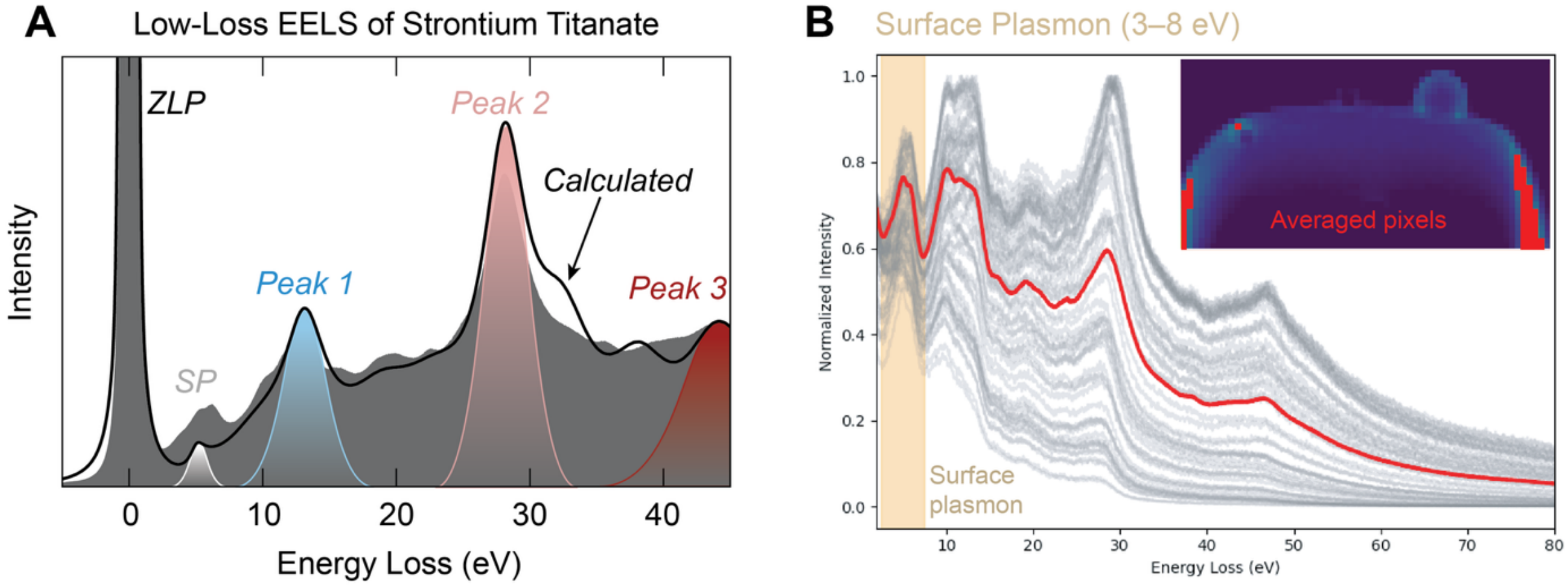

**Fig. S8.**

**Low-loss modes in $SrTiO_3$.** (**A**) The four analyzed low-loss peaks of $SrTiO_3$. (**B**) The average spectrum of the surface plasmon mode is mapped according to averaged STEM-EELS spectra over the surface plasmon region. The intensity in the 3–8 eV window determined the contrast of the inset STEM-EELS map.

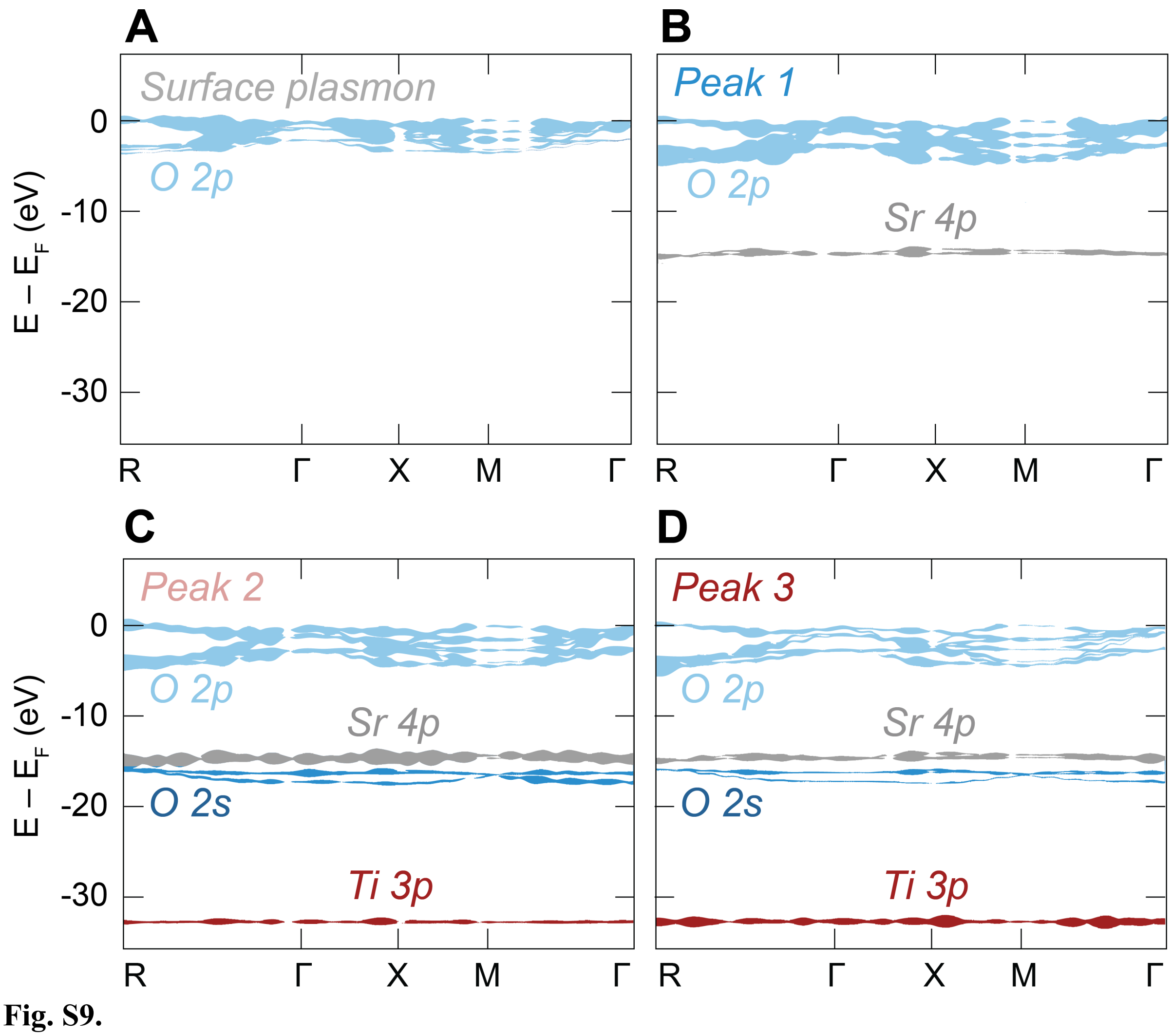

**Fig. S9.**

**TDDFT spectral projections of $SrTiO_3$ low-loss modes.** The peaks indicated in Fig. S8 are spectrally projected for the (**A**) surface plasmon, (**B**) peak 1, (**C**) peak 2, and (**D**) peak 3. TDDFT is implemented for the projection and false colors are applied according to the primary element-specific orbital occupation determined by projected DFT calculations.

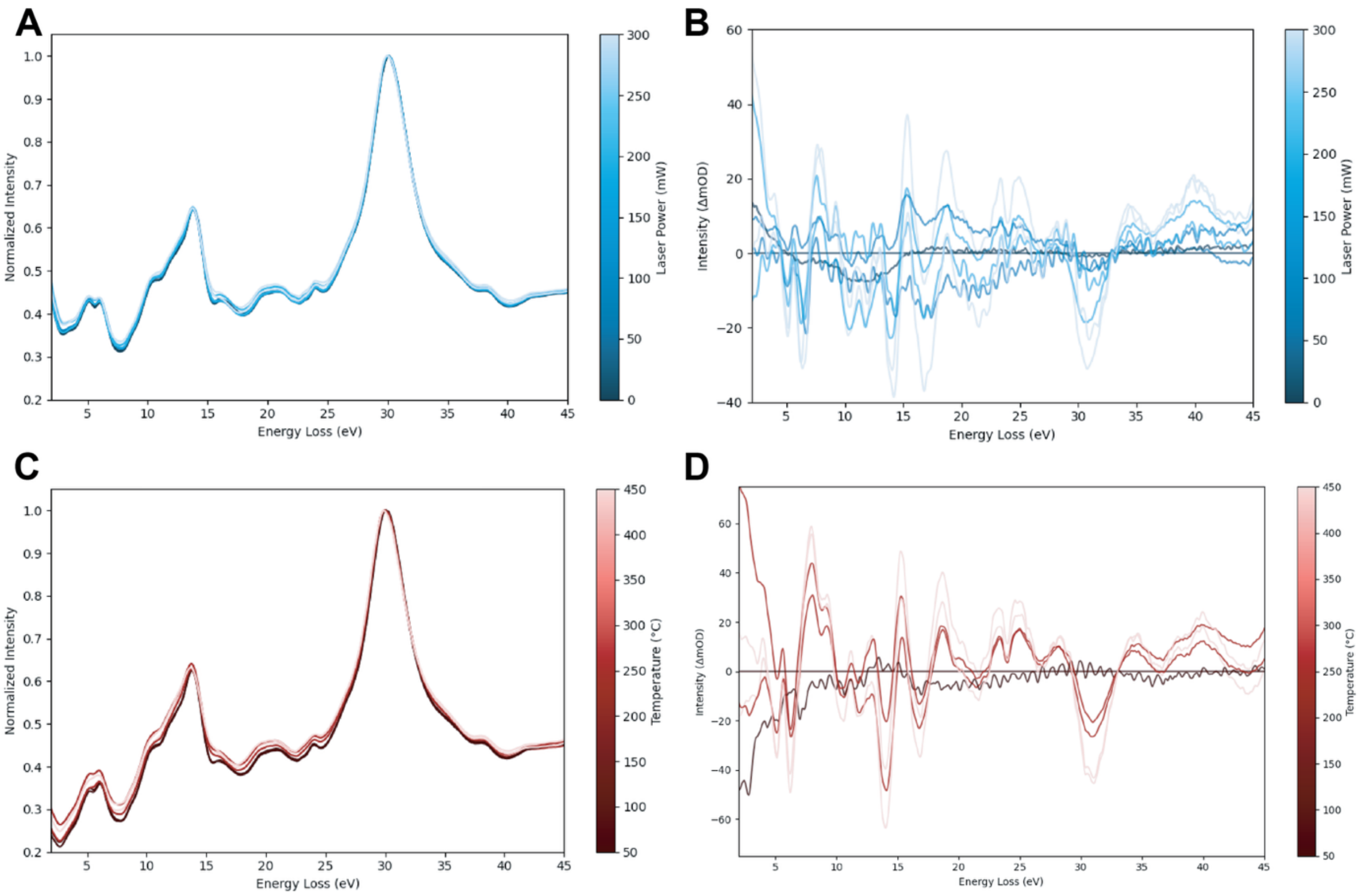

**Fig. S10.**

**Photomodulated and photothermal STEM-EELS hysteresis.** (**A**) All raw image-averaged STEM-EELS hysteresis data for the photomodulated single-particle measurements of $SrTiO_3$. (**B**) Differential spectra more clearly highlight any potential hysteresis where the 0 mW data are acquired sequentially and thus should have no relative hysteresis induced by the laser. (**C**) All raw image-averaged STEM-EELS hysteresis data for the photothermal single-particle measurements on the same region of interest, shown in main text Fig. 4E. (**D**) Differential spectra to emphasize potential hysteresis where again the 50 ˚C trace is measured sequentially after no temperature change and should have no temperature-induced hysteresis.

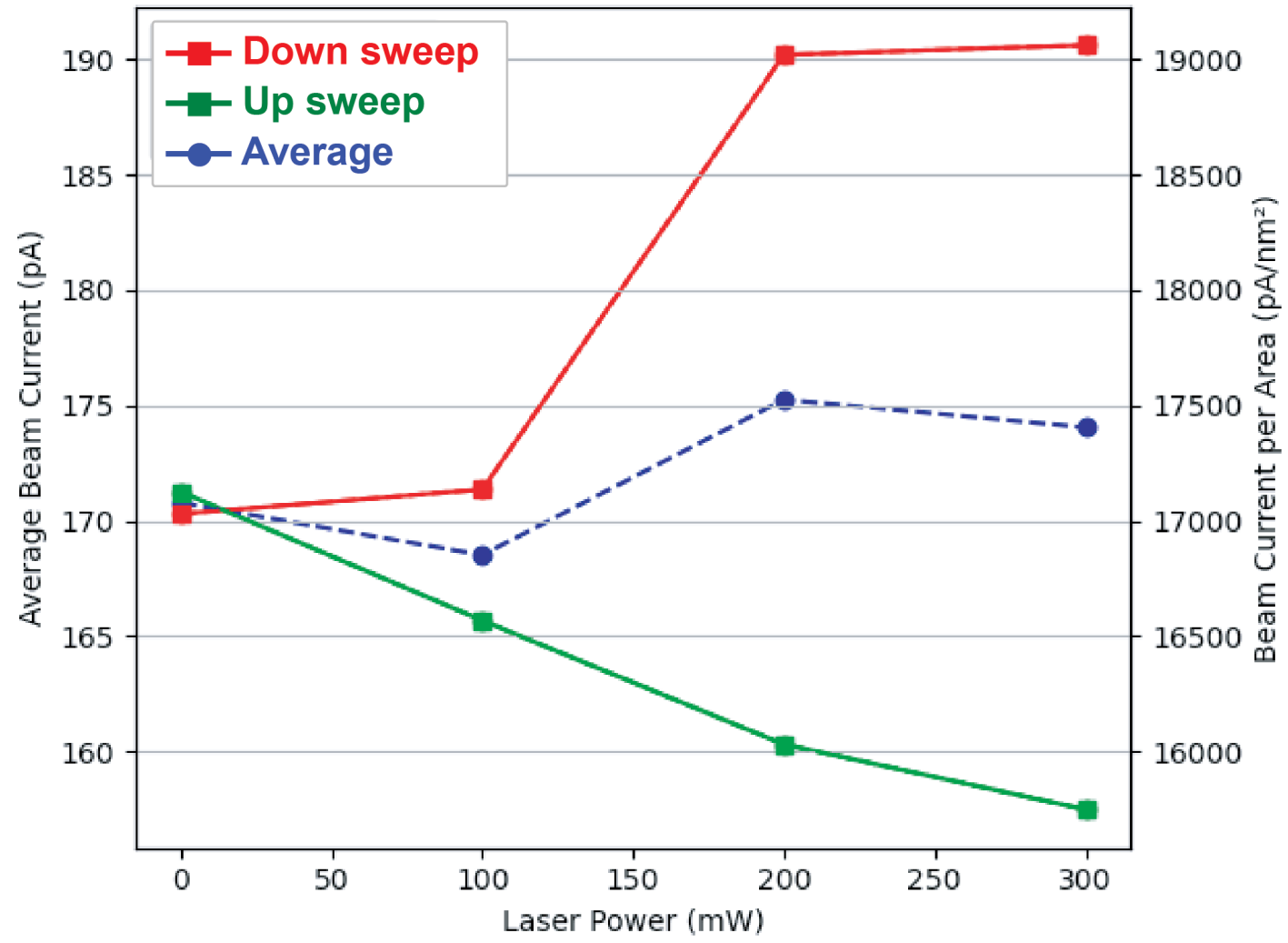

**Fig. S11.**

**Photomodulated STEM-EELS beam current.** Approximate beam current values were recorded for the up and down sweeps as a function of laser power laser power. This approximate beam current was measured using the total counts in each EEL spectrum acquired using the direct electron detector with an additional factor of 25 considered for lost counts due to detector-blocked electrons and high-loss scattering. A 0.25 s × 0.07% live time is considered as the acquisition time with a 1 Å probe size.

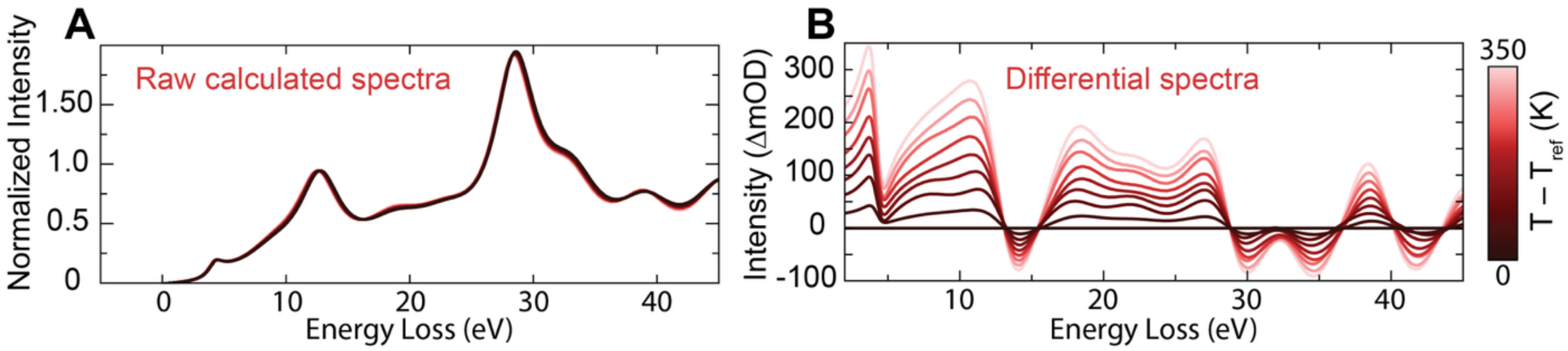

**Fig. S12.**

**Temperature-dependent TDDFT calculations by modeling linear thermal lattice expansion of $SrTiO_3$.** (**A**) Raw temperature-dependent spectra are simulated using TDDFT and an isotropically expanded $SrTiO_3$ unit cell according to its thermal expansion coefficient. (**B**) Simulated differential spectra as a function of lattice temperature due to the thermal expansion.

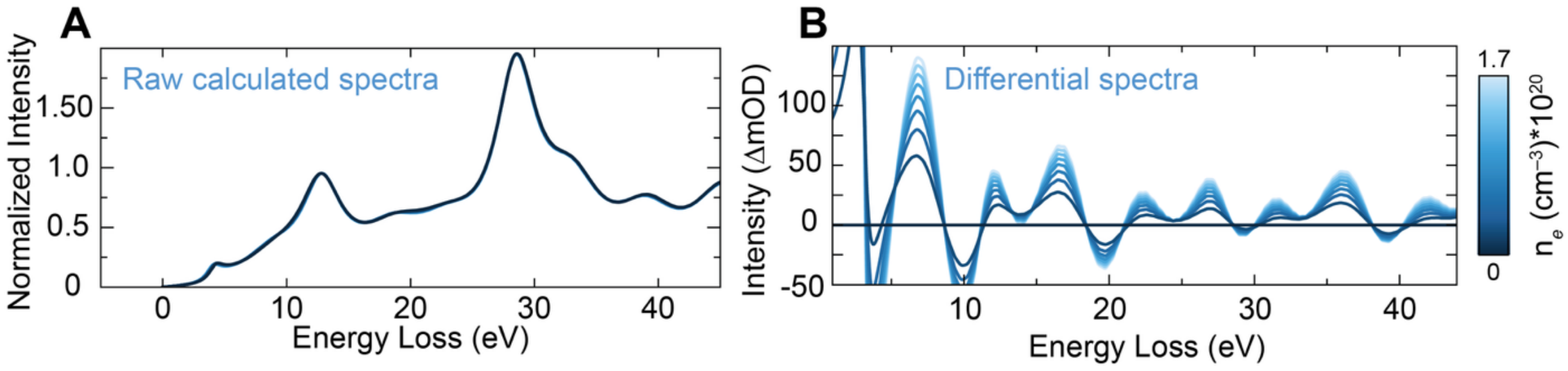

**Figure S13.**

**Photocarrier-dependent TDDFT calculations of $SrTiO_3$.** (**A**) Raw spectra calculated by including two chemical potentials (one for electrons and one for holes), modeled in $SrTiO_3$ to simulate photocarrier-induced spectral effects. (**B**) Corresponding differential spectra to better visualize changes induced by the photocarrier density listed in the color scale.

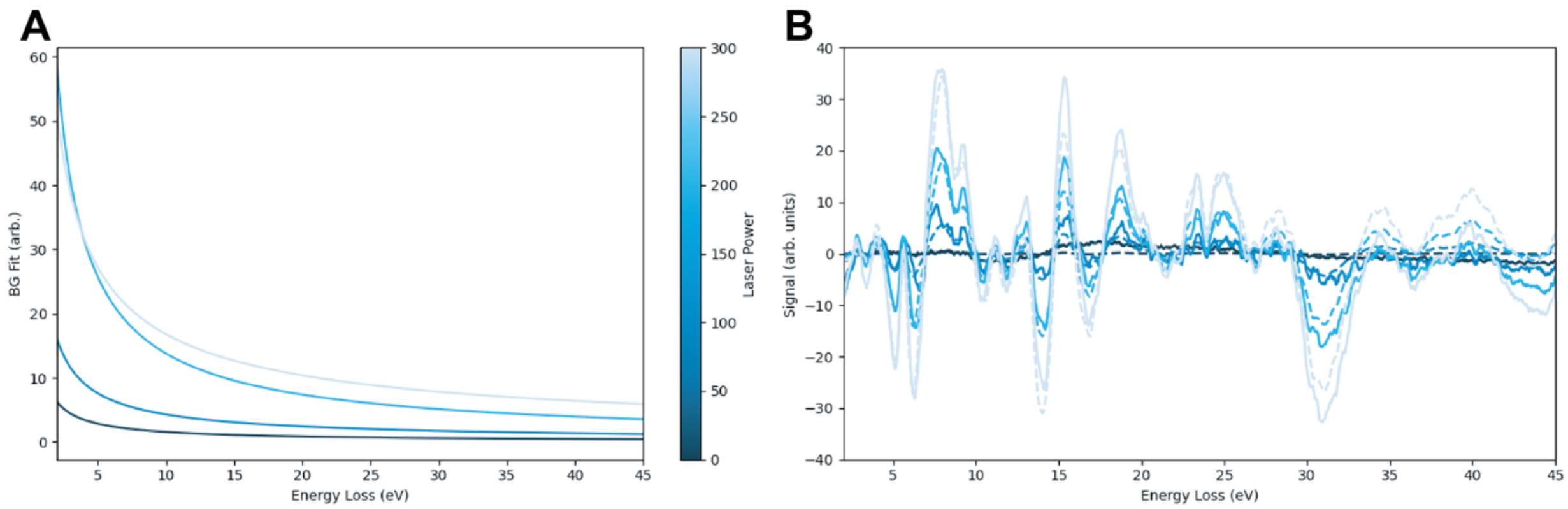

**Figure S14.**

**Background fitting for average photomodulated STEM-EELS datasets.** (**A**) The power-low fitting used to model the fluctuations in the zero-loss peak induced locally by EM field as well as the change in beam current. (**B**) Spectrally fitting the photothermal heating background from the STEM-EELS image-averaged and temperature-dependent dataset to the four average photomodulated STEM-EELS image-averaged datasets. The fittings are uniformly scaled by a factor as shown in (B), and the fit is directly subtracted from these datasets to produce main text Fig. 4C.

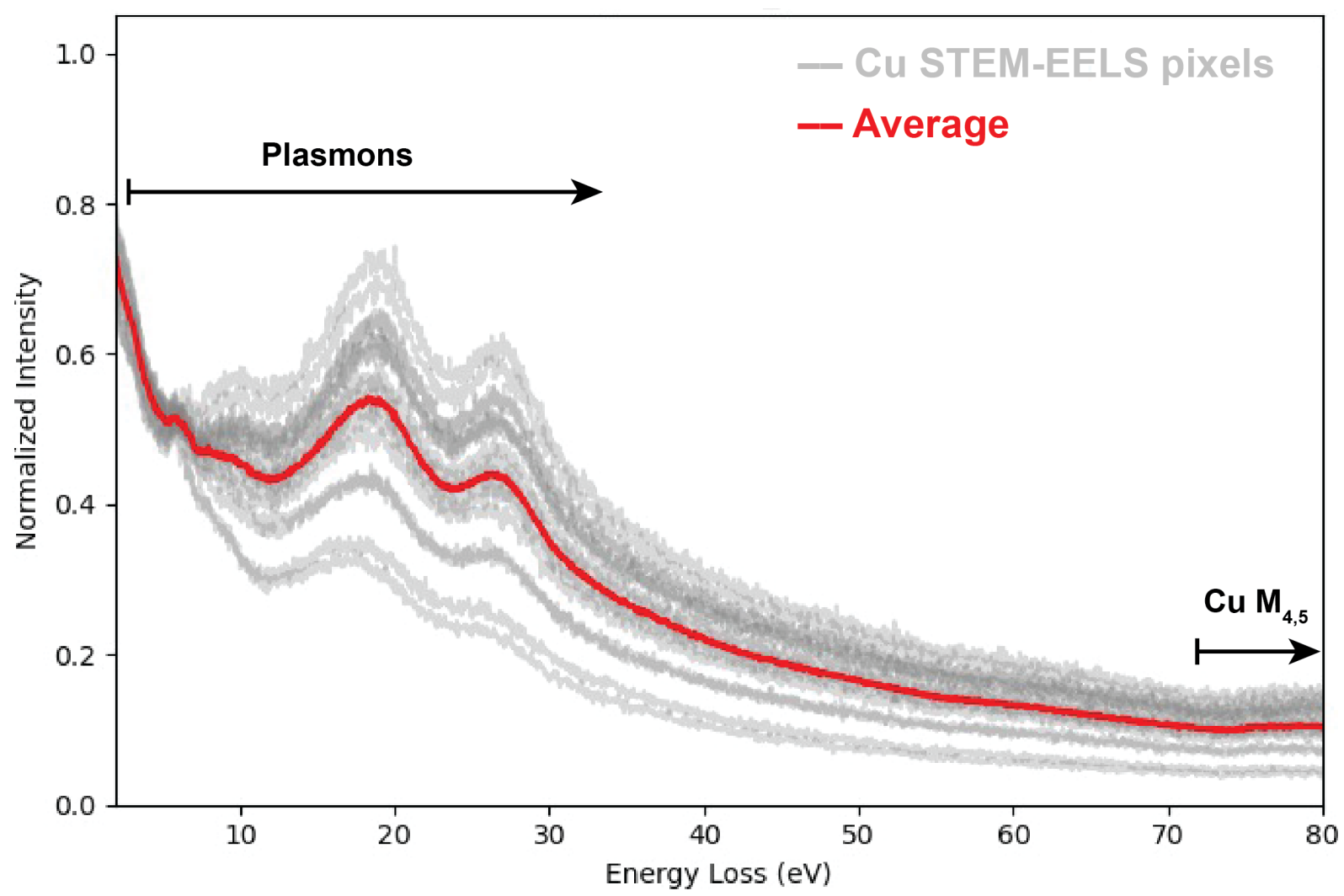

**Fig. S15.**

**STEM-EELS spectra of the Cu nanoparticle's volume plasmon and $M_{4,5}$ edge.** Binning the photomodulated low-loss spectra over the largest Cu nanoparticle's region of interest (shown in Fig. S2C) depicts the ability to resolve the Cu plasmon peaks.

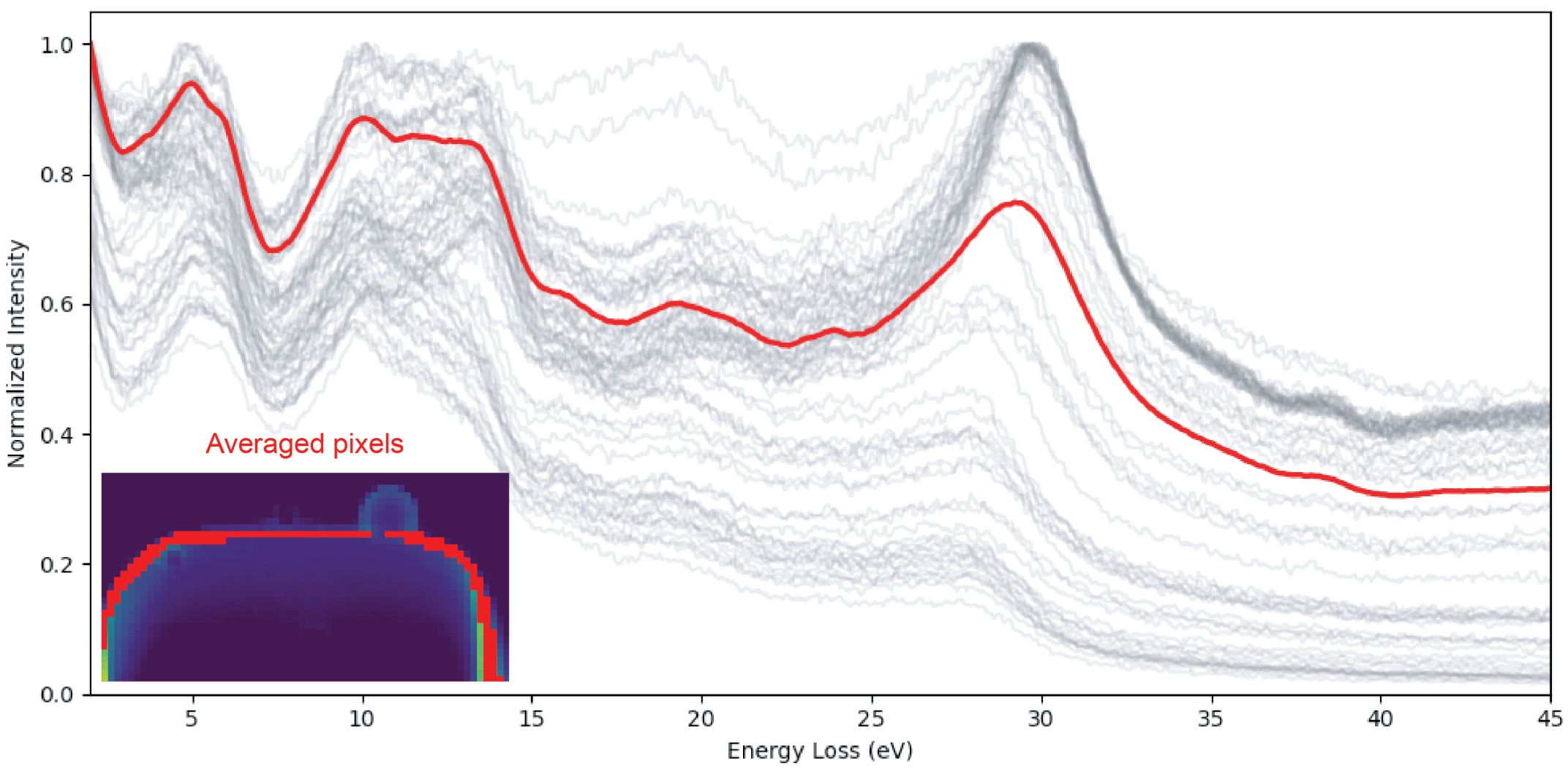

**Figure S16.**

**STEM-EELS spectra over the surface oxygen vacancy regions.** STEM-EELS pixels are averaged over the outer region of the nanoparticle where the oxygen vacancies are localized. The volume plasmon peaks redshift due to the lower carrier density of the disordered oxygen vacancy phase. The inset depicts the surface plasmon intensity STEM-EELS map (3–8 eV) with the pixels averaged highlighted in red.

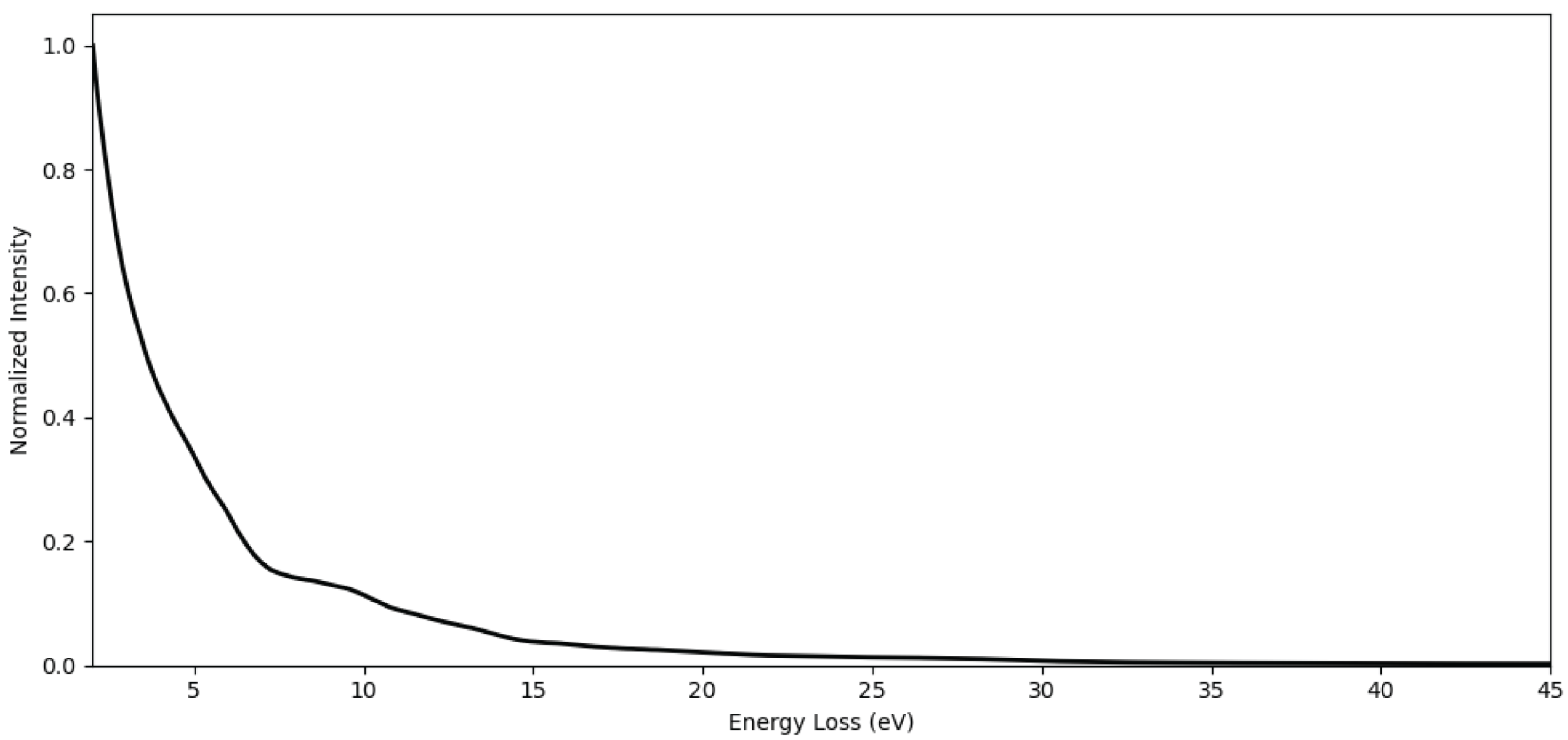

**Figure S17.**

**STEM-EELS spectra over vacuum background regions.** The STEM-EELS-averaged spectrum over vacuum depicts the tail of the ZLP without signal from the nanoparticle. Small features due to aloof scattering may be present.

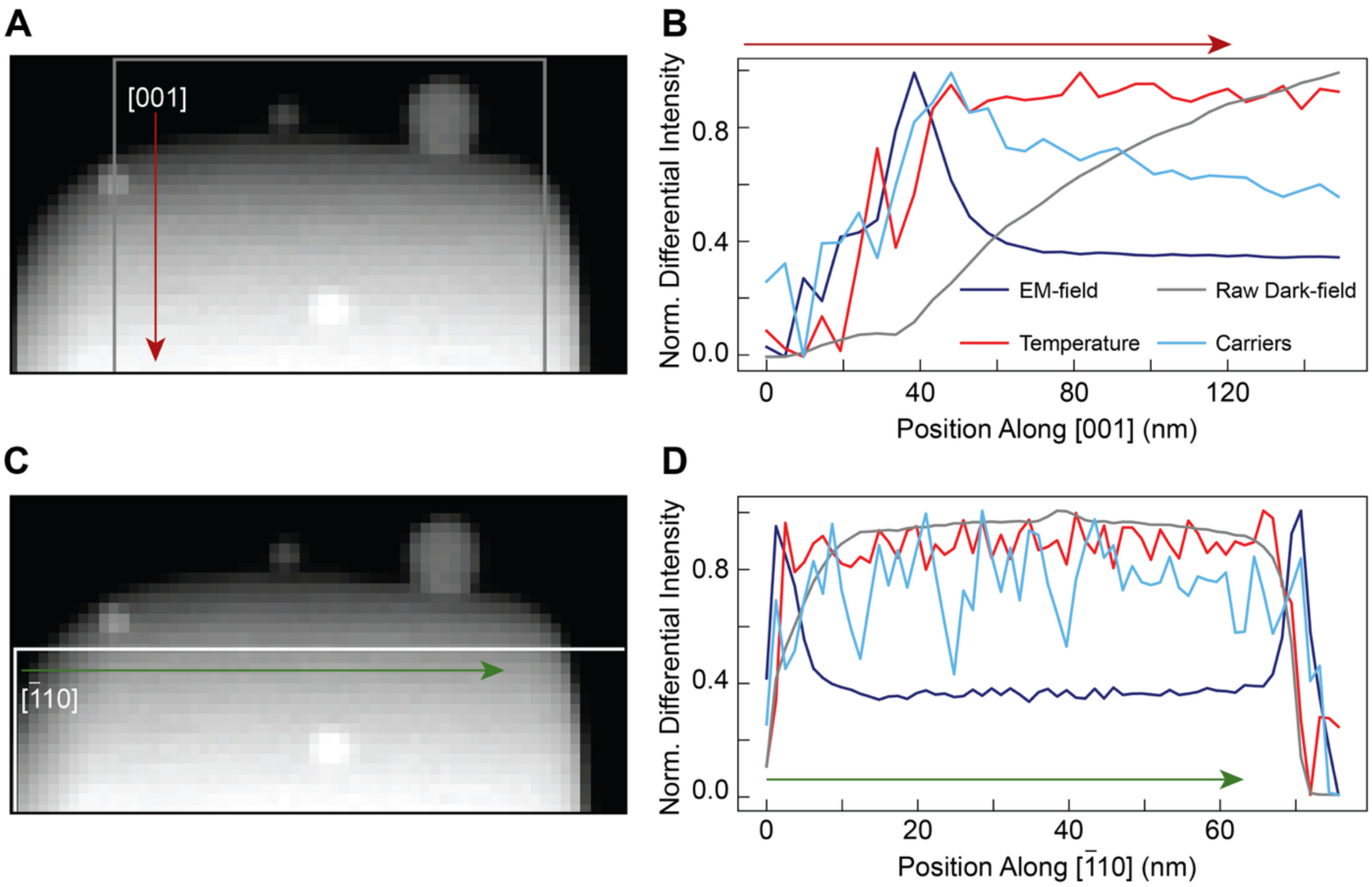

**Figure S18.**

**Comparing facet-averaged differential signals from photomodulated STEM-EELS maps.** (**A**) The average ADF image at 300 mW that highlights the rows being averaged down the nanoparticle along the [001] zone axis. (**B**) The various signals measured with photomodulated STEM-EELS imaging that visualize the EM field strength, photoexcited electron density, photothermal temperature, and dark-field counts (relative thickness) along the [001] zone axis in nanometers. The raw dark-field counts are plotted whereas all other plots reflect the differential intensity. (**C,D**) The dark-field image and differential signal intensities averaged as a function of image column down the $[\bar{1}10]$ zone axis. The analyzed images are depicted in main text Fig. 4E.

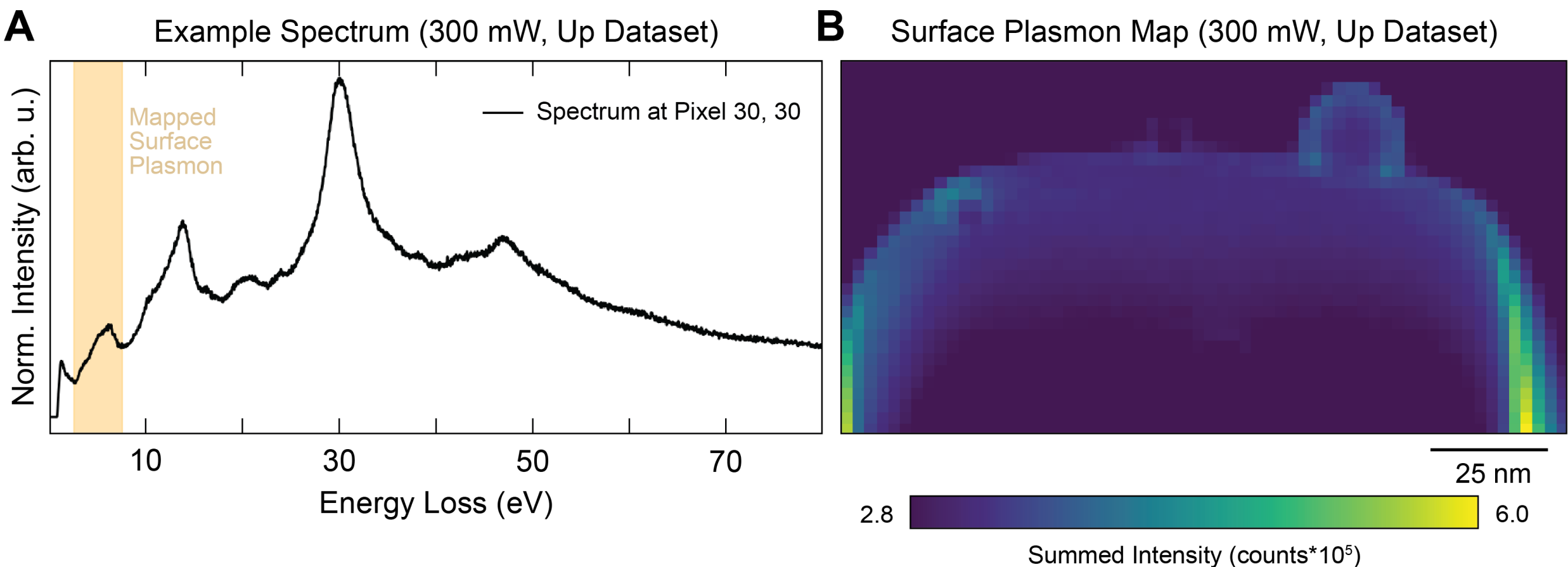

**Fig. S19.**

**STEM-EELS map of the $SrTiO_3$ surface plasmon mode.** (**A**) An example spectrum at pixel 30, 30 in the 300 mW laser set power (last dataset) depicting the mapped surface plasmon from 3–8 eV. (**B**) The STEM-EELS map of the surface plasmon intensity across the nanoparticle.

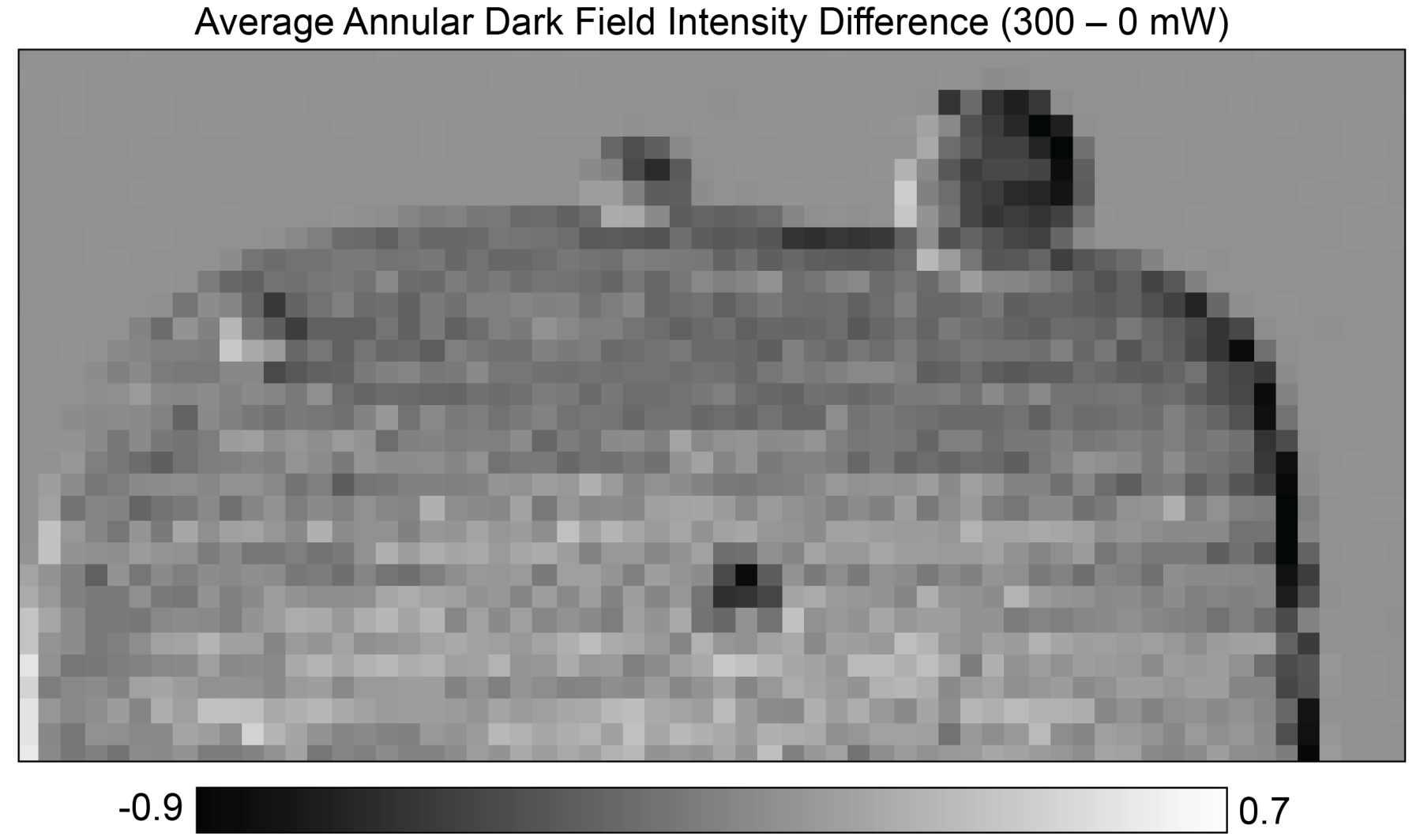

**Fig. S20.**

**Photomodulated image drift analysis.** The dark field intensity difference between the 300 and 0 mW datasets. The dark-field counts for the two 300 mW and 0 mW datasets were separately normalized and averaged prior to calculating the difference.

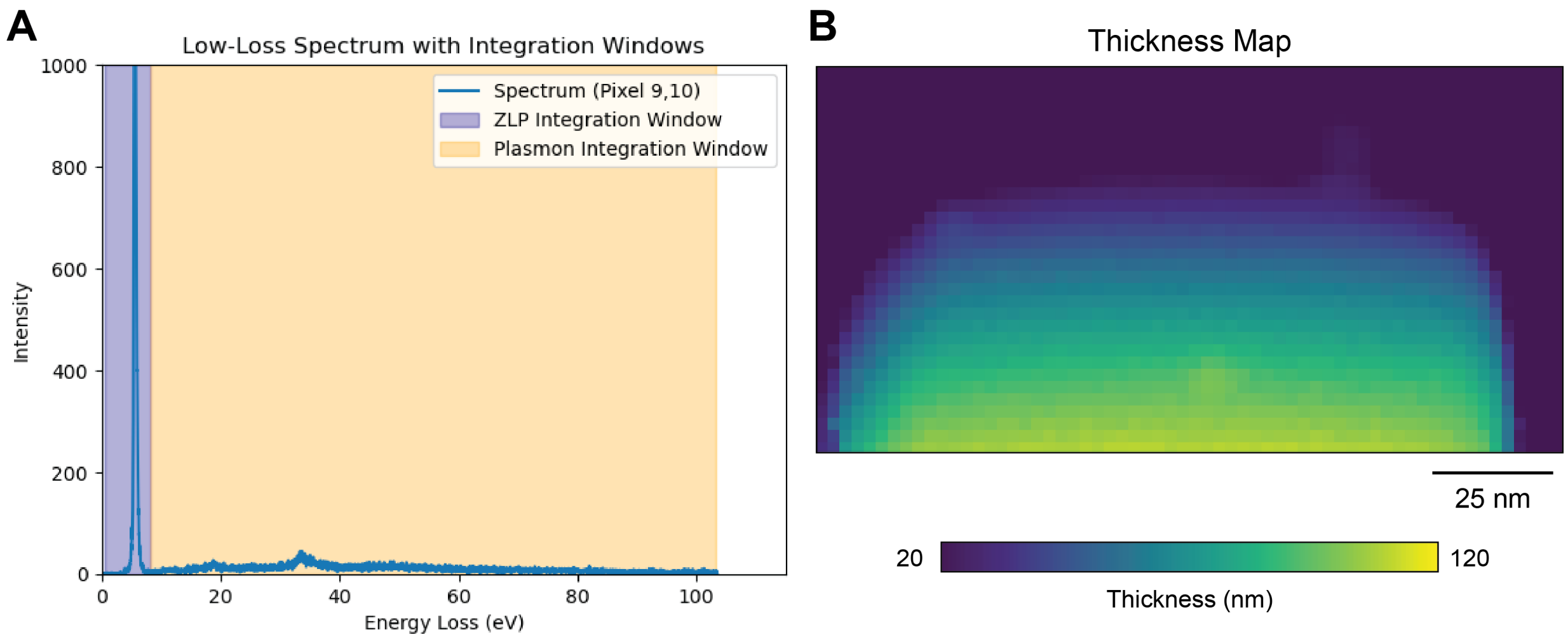

**Fig. S21.**

**Nanoparticle thickness map.** (**A**) An example low-loss spectrum to depict the analysis windows for the log-ratio thickness mapping method. The spectrum at example pixel 9,10 in the STEM-EELS map is plotted for the studied nanoparticle. (**B**) The thickness map evaluated using STEM-EELS and the log-ratio method. A mean free path of 130 nm was approximated for the 200 kV accelerating voltage.

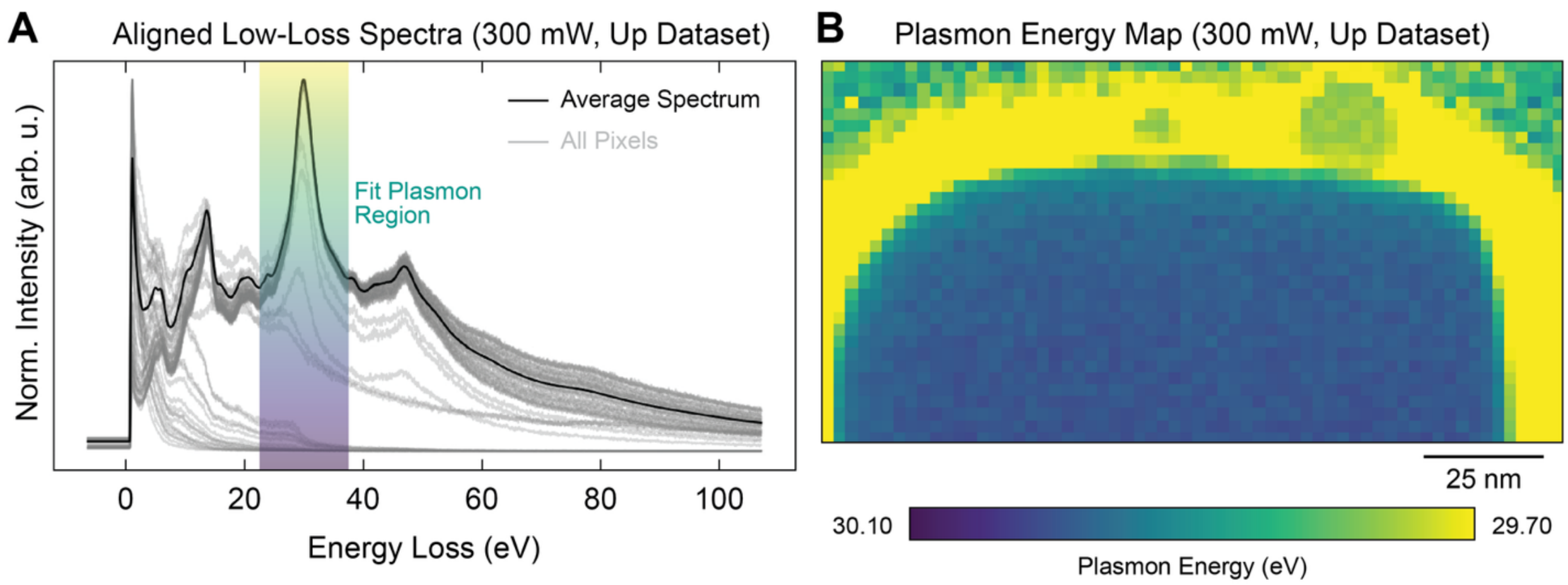

**Fig. S22.**

**Photomodulated STEM-EELS averaging and $SrTiO_3$ bulk plasmon map.** (**A**) The main plasmon analyzed with center-of-mass fitting. The grey spectra are spectra from all pixels in the STEM-EELS image, and the average spectrum is shown in black. (**B**) A map of the fit plasmon energy across the STEM-EELS map, where only the image pixels over $SrTiO_3$ are quantitative. This result highlights the redshift of the plasmon energy at the nanoparticle's surface.

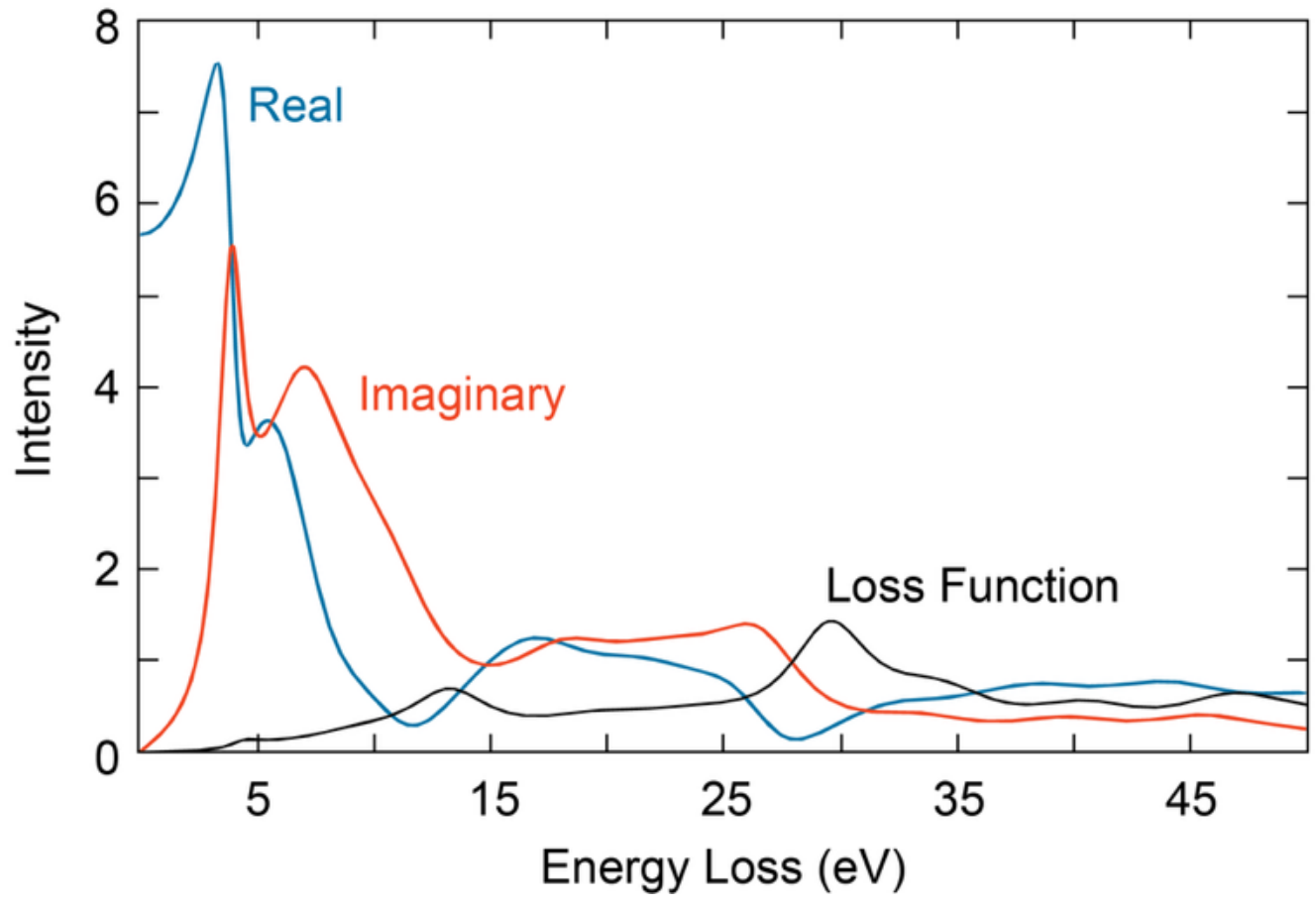

**Fig. S23.**

**Calculated dielectric function of $SrTiO_3$.** The real and imaginary components of the complex dielectric function of undoped strontium titanate are calculated as a function of energy using TDDFT in turboEELS. The calculated loss function is included for reference.

**Table S1.**

**Photomodulated STEM-EELS laser irradiance values and hysteresis cycle.** The set laser power and approximate irradiance value in the order used for the hysteresis cycling. Corrected irradiance accounts for photons lost due to fiber coupling, the TEM optical window, and reflections. Approximately 35% of the laser's set beam irradiance (column 2) is illuminates the specimen after losses.

| **Laser Power (mW)** | **Irradiance ($mW\cdot cm^{-2}$)** | **Irradiance Adjusted for Losses ($mW\cdot cm^{-2}$)** | **Acquisition Order** |
|---|---|---|---|
| 300 | $1.5\times10^{7}$ | $5.3\times10^{6}$ | 1 |
| 200 | $1.0\times10^{7}$ | $3.6\times10^{6}$ | 2 |
| 100 | $0.51\times10^{7}$ | $1.8\times10^{6}$ | 3 |
| 0 | 0 | 0 | 4 |
| 100 | $0.51\times10^{7}$ | $1.8\times10^{6}$ | 5 |
| 200 | $1.0\times10^{7}$ | $3.6\times10^{6}$ | 6 |
| 300 | $1.5\times10^{7}$ | $5.3\times10^{6}$ | 7 |

**Table S2.**

**Thermal control STEM-EELS temperature values and hysteresis cycle.** Set temperature during the STEM-EELS hysteresis cycles in order of acquisition.

| **Temperature (˚C)** | **Acquisition Order** |
|---|---|
| 450 | 1 |
| 250 | 2 |
| 50 | 3 |
| 250 | 4 |
| 450 | 5 |

**Movie S1.**

**Spectral projection across all energies of the $SrTiO_3$ low-loss spectrum.**

**Movie S2.**

**Nanoparticle position over all photomodulated measurements.**

**Movie S3.**

**Average power-dependent field map**

**Movie S4.**

**Average power-dependent carrier map**

**Movie S5.**

**Average power-dependent temperature map**

**Movie S6.**

**Average power-dependent dark-field intensity differential map**